\documentclass[5p,times]{elsarticle}
\usepackage[utf8]{inputenc}
\usepackage{acro}
\usepackage{makecell}
\usepackage{todonotes}
\usepackage[caption=false]{subfig}
\usepackage{listings}
\usepackage{graphicx}
\usepackage{url}
\usepackage{booktabs}
\usepackage{makecell}
\usepackage{amssymb}
\usepackage{placeins}
\usepackage{float}
\usepackage{caption}
\usepackage[english]{babel}
\usepackage{chngcntr}

\let\xtodo\todo
\renewcommand{\todo}[1]{\xtodo[inline,color=orange!75]{#1}}

\DeclareAcronym{IV}{short = IV, long = independent variable}
\DeclareAcronym{DV}{short = DV, long = dependent variable}
\DeclareAcronym{UI}{short = UI, long = user interface}
\DeclareAcronym{TLX}{short = NASA-TLX, long = NASA Task Load Index}
\DeclareAcronym{RTLX}{short = raw TLX, long =raw NASA-Task Load Index}
\DeclareAcronym{ER}{short = ER, long = error rate}
\DeclareAcronym{TCT}{short = TCT, long = Task Completion Time}
\DeclareAcronym{HCI}{short = HCI, long = Human-Computer Interaction}
\DeclareAcronym{HFE}{short = HFE, long = Human Factors and Ergonomics}
\DeclareAcronym{cuDNN}{short = cuDNN, long =  CUDA Deep Neural Network library}
\DeclareAcronym{RMSE}{short = RMSE, long = root mean squared error}
\DeclareAcronym{HMD}{short = HMD, long = Head-Mounted Display}
\DeclareAcronym{RF}{short = RF, long = Random Forest}
\DeclareAcronym{GP}{short = GP, long = Gaussian process, long-plural = Gaussian processes}
\DeclareAcronym{KNN}{short = \textit{k}NN, long = \textit{k}-nearest neighbor}
\DeclareAcronym{NN}{short = NN, long = Neural Network}
\DeclareAcronym{DNN}{short = DNN, long =  Deep Neural Network}
\DeclareAcronym{CNN}{short = CNN, long = Convolutional Neural Network}
\DeclareAcronym{FCL}{short = FCL, long = fully connected layer}
\DeclareAcronym{BoD}{short = BoD, long = Back-of-Device}
\DeclareAcronym{VR}{short = VR, long = Virtual Reality}
\DeclareAcronym{AR}{short = AR, long = Augmented Reality}
\DeclareAcronym{MR}{short = MR, long = Mixed Reality}
\DeclareAcronym{FOV}{short = FoV, long = field of view}
\DeclareAcronym{RW}{short = RW, long = real world}
\DeclareAcronym{IFRC}{short = IFRC, long = index finger ray cast}
\DeclareAcronym{FARC}{short = FARC, long = forearm ray cast}
\DeclareAcronym{EFRC}{short = EFRC, long = eye-finger ray cast}
\DeclareAcronym{HRC}{short = HRC, long = head ray cast}
\DeclareAcronym{DOF}{short = DOF, long = degree-of-freedom, long-plural-form = degrees-of-freedom}
\DeclareAcronym{6DOF}{short = 6DOF, long = six-degree-of-freedom}
\DeclareAcronym{3DOF}{short = 3DOF, long = three-degree-of-freedom}
\DeclareAcronym{LOOCV}{short = LOOCV, long = leave-one-out cross-validation}
\DeclareAcronym{LOPOCV}{short = LOPOCV, long = leave-one-participant-out cross-validation}
\DeclareAcronym{CV}{short = CV, long = cross-validation}
\DeclareAcronym{RM}{short = RM, long = repeated measure}
\DeclareAcronym{ANOVA}{short = ANOVA, long = analysis of variance}
\DeclareAcronym{RMANOVA}{short = RM-ANOVA, long = repeated measures analysis of variance}
\DeclareAcronym{MANOVA}{short = MANOVA, long = multivariate analysis of variance}
\DeclareAcronym{AGATe}{short = AGATe, long = AGreement Analysis Toolkit}
\DeclareAcronym{GHoST}{short = GHoST, long = Gesture Heatmap Toolkit Gesture Heatmaps Toolkit}
\DeclareAcronym{GREAT}{short = GREAT, long = Gesture Relative Accuracy Toolkit}
\DeclareAcronym{GRT}{short = GRT, long = Gesture Recognition Toolkit}
\DeclareAcronym{DTW}{short = DTW, long = Dynamic Time Warping}
\DeclareAcronym{LHRD}{short = LHRD, long = large high resolution display}
\DeclareAcronym{GEQ}{short = GEQ, long = Game Experience Questionnaire}
\DeclareAcronym{SPGQ}{short = SPGQ, long = Social Presence Gaming Questionnaire}
\DeclareAcronym{SUS}{short = SUS, long = Slater-Usoh-Steed questionnaire}
\DeclareAcronym{IPQ}{short = IPQ, long = igroup presence questionnaire}
\DeclareAcronym{PQ}{short = WS, long = Witmer and Singer presence questionnaire}
\DeclareAcronym{ITQ}{short = ITQ, long = immmersive tendency questionnaire}
\DeclareAcronym{BIP}{short = BIP, long = break-in-presence}
\DeclareAcronym{VE}{short = VE, long = virtual environment}
\DeclareAcronym{TPI}{short = TPI, long = temple presence inventory}
\DeclareAcronym{ITC}{short = ITC-SOPI, long = ITC-Sense of presence inventory}
\DeclareAcronym{IMU}{short = IMU, long = inertial measurement unit}
\DeclareAcronym{GUI}{short = GUI, long = graphical user interface}
\DeclareAcronym{TAM}{short = TAM, long = Technology Acceptance Model}
\DeclareAcronym{UTAUT}{short = UTAUT, long = Unified Theory of Acceptance and Use of Technology}
\DeclareAcronym{AI}{short = AI, long = Artificial Intelligence}
\DeclareAcronym{DALI}{short = DALI, long = Driving Activity Load Index}
\DeclareAcronym{BLMM}{short = BLMM, long = Bayesian linear mixed model}
\DeclareAcronym{HCAI}{short = HCAI, long = Human-Centered Artificial Intelligence}
\DeclareAcronym{IUI}{short = IUI, long = Intelligent User Interfaces}
\DeclareAcronym{EDA}{short = EDA, long = Electrodermal Activity}
\DeclareAcronym{EEG}{short = EEG, long = Electroencephalography}
\DeclareAcronym{API}{short = API, long = Application Programming Interface}
\DeclareAcronym{LLM}{short = LLM, long = Large Language Model}
\DeclareAcronym{EUP}{short = EUP, long = End-User Programming}

\newcommand\sbullet[1][.75]{\mathbin{\vcenter{\hbox{\scalebox{#1}{$\bullet$}}}}}

\newcommand{\lastaccess}{last access 20-10-2023}

\biboptions{longnamesfirst,comma}

\title{Large Language Models to the Rescue: Reducing the Complexity in Scientific Workflow Development Using ChatGPT}

\author[1]{Mario Sänger\corref{cor1}}
\ead{saengema@informatik.hu-berlin.de}

\author[1]{Ninon De Mecquenem}
\ead{mecquenn@informatik.hu-berlin.de}

\author[2,3]{Katarzyna Ewa Lewińska}
\ead{lewinska@geo.hu-berlin.de}

\author[1]{Vasilis Bountris}
\ead{vasilis.bountris@informatik.hu-berlin.de}

\author[1]{Fabian Lehmann}
\ead{fabian.lehmann@informatik.hu-berlin.de}

\author[1]{Ulf Leser\corref{cor1}}
\ead{leser@informatik.hu-berlin.de}

\author[1]{Thomas~Kosch\corref{cor1}}
\ead{thomas.kosch@hu-berlin.de}

\cortext[cor1]{Corresponding authors}

\affiliation[1]{organization={Humboldt Universit\unexpanded{ä}t zu Berlin, Department of Computer Science},
addressline={Unter den Linden 6},
city={Berlin},
postcode={10099},
country={Germany}}

\affiliation[2]{organization={Humboldt Universit\unexpanded{ä}t zu Berlin, Department of Geography},
addressline={Unter den Linden 6},
city={Berlin},
postcode={10099},
country={Germany}}

\affiliation[3]{organization={University of Wisconsin-Madison, Department of Forest and Wildlife Ecology},
addressline={1630 Linden Drive},
city={Madison, WI},
postcode={53706},
country={United States}}

\begin{document}
\begin{abstract}
Scientific workflow systems are increasingly popular for expressing and executing complex data analysis pipelines over large datasets, as they offer reproducibility, dependability, and scalability of analyses by automatic parallelization on large compute clusters. However, implementing workflows is difficult due to the involvement of many black-box tools and the deep infrastructure stack necessary for their execution. Simultaneously, user-supporting tools are rare, and the number of available examples is much lower than in classical programming languages. To address these challenges, we investigate the efficiency of Large Language Models (LLMs), specifically ChatGPT,  to support users when dealing with scientific workflows. We performed three user studies in two scientific domains to evaluate ChatGPT for comprehending, adapting, and extending workflows. Our results indicate that LLMs efficiently interpret workflows but achieve lower performance for exchanging components or purposeful workflow extensions. We characterize their limitations in these challenging scenarios and suggest future research directions.  
\end{abstract}




\maketitle

\section{Introduction}

Large-scale data analysis pipelines (also known as scientific workflows) are crucial in driving research advances for natural sciences~\cite{10.1145/1376616.1376772}. They are pivotal in accelerating large and complex data analysis on distributed infrastructures and offer essential features, such as reproducibility and dependability~\cite{COHENBOULAKIA2017284}. In bioinformatics, for instance, scientific workflows are analyzing the terabyte-large data sets produced by modern DNA or RNA sequencing machines in a wide variety of experiments~\cite{WWG21}, thereby aiding in building a comprehensive understanding of biological processes and human diseases. Bioinformatics workflows typically include many individual computational steps, such as data pre-processing, extensive quality control, aggregation of raw sequencing data into consensus sequences, machine-learning-based tasks for classification and clustering, statistical assessments, and result visualization. Each step is carried out by a specific program, typically not written by the workflow developer but exchanged within a worldwide community of researchers~\cite{IIC19}. Execution of a workflow on a distributed infrastructure, in principle, is taken care of by a workflow engine; however, the idiosyncrasies of the different infrastructures (e.g., file system, number and features of compute nodes, applied resource manager, and scheduler) often require workflow users to tune their scripts individually for every new system~\cite{FSCC+21}.

However, typical developers of workflows are researchers from heterogeneous scientific fields who possess expertise in their respective domains but often lack in-depth knowledge in software development or distributed computing. They often encounter difficulties understanding the complex implementations of exchanged codes and the deep infrastructure stack necessary for their distributed execution. This situation challenges efficient workflow implementation, slows down or hinders data exploration and scientific innovation processes~\cite{cohen2011search}. Consequently, low human productivity is a significant bottleneck in the creation, adaption, and interpretation of scientific workflows~\cite{DPA+18}.
%

%

Parallel to this, there is a well-established field in human-computer interaction focusing on assisting end-user programmers and software development, as highlighted by previous work \cite{myers1992survey, barricelli2019end, lau2021tweakit, 10.1145/3540250.3549084}, to reduce the perceived cognitive workload and improve the overall programmer performance~\cite{kosch2023a}.
Research in this field includes programming-by-demonstration \cite{li2017sugilite,sereshkeh2020vasta}, visual programming \cite{tamilselvam2019visual,coronado2021towards},  and natural language instructions \cite{li2019end}. 
Recent work in this area particularly investigated prospects of general-purpose \acp{LLM}, such as ChatGPT~\cite{ouyang2022training}, LLaMA~\cite{touvron2023llama} and Bloom\cite{scao2022bloom}, for supporting end-user-programming \cite{bimbatti2023can,sobania2023analysis,liu2023wants} and software development in general~\cite{white2023chatgpt,surameery2023use}. 
For instance, Bimbatti et al. \cite{bimbatti2023can} explore using ChatGPT to enhance natural language understanding within an end-user development environment, assisting non-expert users in developing programs for collaborative robots. 
Moreover, White et al.~\cite{white2023chatgpt} introduce prompt design techniques for automating typical software engineering tasks, including ensuring code independence from third-party libraries and generating an API specification from a list of requirements. 
Surameery et al.~\cite{surameery2023use} evaluate LLMs for supporting code debugging. 
However, results from such studies, which focus on standard programming languages, cannot easily be transferred to workflow systems. Workflow scripts mostly call external tools with agnostic names and have little recognizable control structures or protected keywords. Publicly available examples are scarce; for instance, the community repository of the popular workflow system Nextflow~\cite{EPF+20} currently offers only 55 released workflows\footnote{\url{https://nf-co.re/stats} - \lastaccess}. Furthermore, workflows can only be understood when the distributed system underlying their execution is considered, creating dependencies much different than usual programs. 
Moreover, studies investigating how LLMs support users in data science - the field in which workflows are applied extensively - do not address the unique characteristics of scientific workflows either and are limited to theoretical considerations~\cite{almarie2023use, hassani2023role}. Practical studies, especially those involving real users, are badly missing.
%

In this work, we address these shortcomings by describing three user-studies in two different scientific fields (biomedicine and Earth observation) that evaluate the suitability of ChatGPT for comprehending, modifying, and extending scientific workflows. Specifically, we evaluate the correctness of ChatGPT regarding explainability, exchange of software components, and extension when providing real-world scientific workflow scripts. 
Our results show a high accuracy for comprehending and explaining scientific workflows but reduced performance for modifying and extending workflow scripts. The domain experts positively assessed the explainability in qualitative inquiries, emphasizing the time-saving capabilities of using \acp{LLM} while engineering existing workflows. Overall, our work indicates that general-purpose LLMs have the potential to improve human performance when analyzing complex scientific workflows.

\section{Related Work}

Previous research investigated how related domains, such as programming, can be augmented using interactive technologies~\cite{10.1145/3540250.3549084, 10.1145/3205873.3210702}. In contrast to programming, where applications use a single programming language and are often executed on a single system, scientific workflows combine multiple software artifacts on distributed stacks for advanced data processing. We ground the reader by providing a literature review about scientific workflows and introducing large language models, including their utility to facilitate the creation of software artifacts.

\subsection{Scientific Workflows}
\begin{figure*}
    \centering
    \subfloat[][]{
        \includegraphics[trim={0 0 285mm 0},clip=true,height=0.5\textwidth]{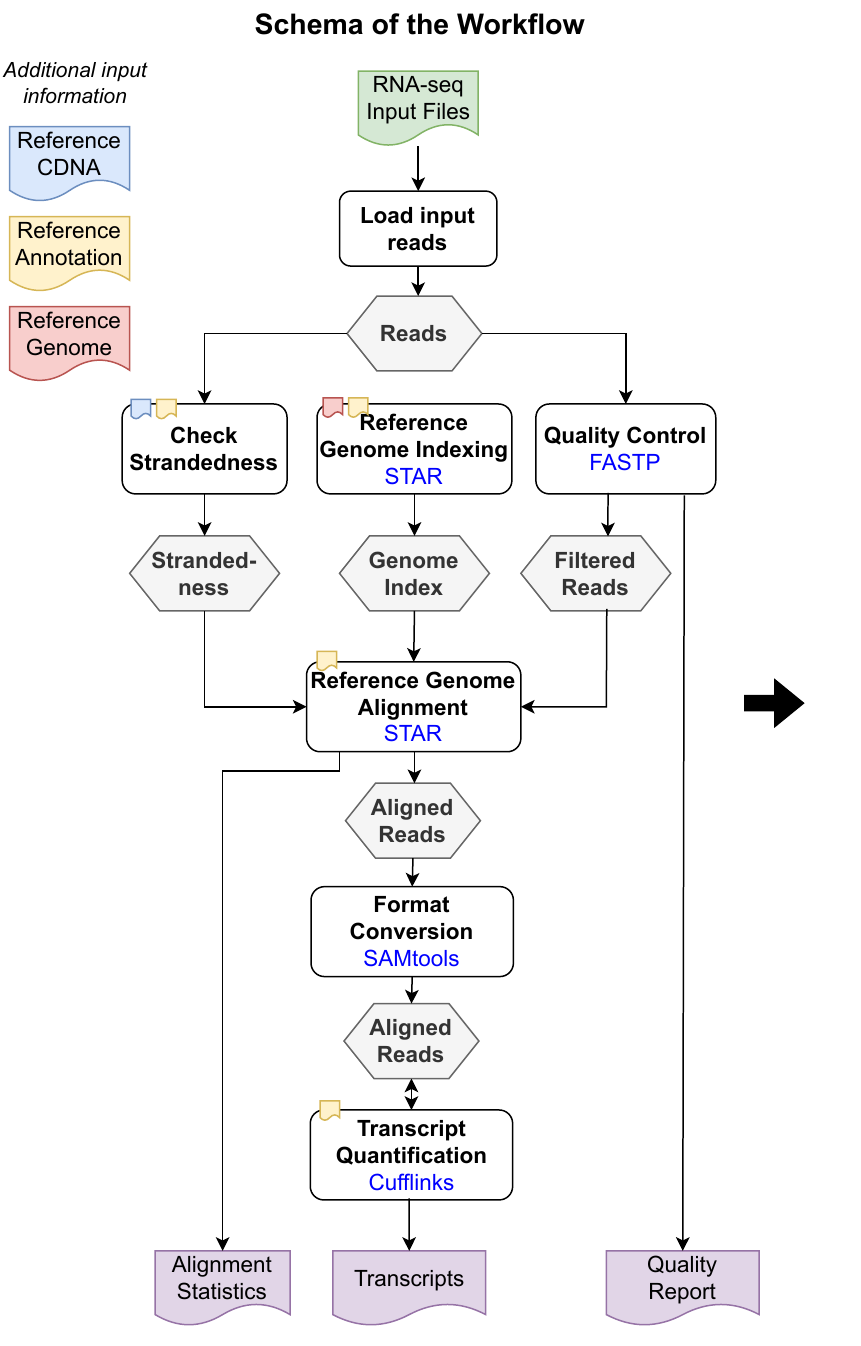}
        \label{fig:ninon_a}
    }
    \subfloat[][]{
            \includegraphics[trim={140mm 0 139mm 0},clip=true,height=0.5\textwidth]{images/workflow-ninon.pdf}
        \label{fig:ninon_b}
    }
    \subfloat[][]{
            \includegraphics[trim={292mm 0 0 0},clip=true,height=0.5\textwidth]{images/workflow-ninon.pdf}
        \label{fig:ninon_c}
    }
    \caption{Example bioinformatics workflow for differential gene expression analysis created by a domain expert recruited in our study. The figure highlights the conceptual schema of the workflow (a), its implementation in Nextflow (b), and the implementation of one single step (c), i.e., reference genome alignment using the STAR tool. The workflow comprises six computational steps in total. For each step the used tool is given in blue below the task name.
    }
    \label{fig:ninon}
\end{figure*}

Scientific workflows are widely used by diverse research communities, such as biomedicine \cite{wratten2021reproducible}, astronomy \cite{ahmad2022efficient}, climatology \cite{kunkel2020potential}, and Earth observation \cite{lehmann2021force} to manage the dataflow and distributed execution of complex analyses, simulations, and experiments.
A scientific workflow comprises a series of interconnected computational steps, often with diverse patterns of dependencies, that define how to process and analyze data to reach a particular research objective. 
Scientific workflows can be regarded as directed acyclic graphs in which the nodes represent computational tasks or operations and edges model dependencies or dataflow between these tasks. An edge from one node to another signifies that the output of the first task is used as input for the second \cite{yu2005taxonomy}.
For example, Figure~\ref{fig:ninon_a} illustrates the computational steps, the tools, and the data flow of a bioinformatics workflow for performing differential gene expression analysis. 
%
%
During workflow execution, a single computation step often involves multiple processes, which are typically executed in a distributed fashion on different machines and batches of the input data, resulting in a much more complex execution graph.
Consequently, scientific workflows help facilitate the reproducibility and traceability of data analyses by explicitly outlining the steps and parameters involved \cite{liew2016scientific}. 
Furthermore, they allow for automation, scaling, and optimization of computational processes, which is especially critical in disciplines dealing with large datasets \cite{gil2007examining}.
The increasing importance of scientific workflows for scientific progress has led to a growing interest in developing more user-friendly tools and methods through the research community.
Scientific workflow management systems, like Apache Airflow \cite{harenslak2021data}, Galaxy \cite{goecks2010galaxy}, Nextflow \cite{di2017nextflow}, Pegasus \cite{deelman2015pegasus}, and Snakemake \cite{koster2012snakemake}, are specifically developed to support users in designing and executing scientific workflows in various aspects.
Key features of such management systems typically include workflow design and composition, (distributed) workflow execution and scheduling, provenance tracking, recovery and failure handling, and resource management \cite{liew2016scientific}. 
Figure~\ref{fig:ninon_b} highlights the implementation of the example workflow as well as a single computational step (see Figure~\ref{fig:ninon_c}), i.e., reference genome alignment using the STAR toolkit in Nextflow.  

Scientific workflows are often reused and adopted for complex data analysis \cite{cohen2011search}. Most of the time, users of scientific workflows are unaware of the workflow's internal functionality and technical details. Instead, users of scientific workflows represent domain experts, such as mathematicians, physicians, or bioinformatics, who are experts in their respective domains but not necessarily in programming and interpreting scientific workflows. Existing scientific workflows were implemented and maintained by persons other than the domain user. This reduces the direct interaction with scientific workflows to a minimum, where domain experts only hand in the input data and evaluate the output data. 
%
Consequently, domain experts using a workflow often do not have the knowledge to modify, extend, or interpret the details of scientific workflows.

\subsection{Large Language Models}
Language Models, such as BERT \cite{kenton2019bert}, GPT-3 \cite{brown2020language},  Bloom \cite{scao2022bloom} and PaLM-2 \cite{anil2023palm}, build the foundation of many recent advancements in natural language processing and understanding.
These models have billions of parameters and are generally pre-trained on vast sets of texts from the web and other repositories, enabling them to encode syntactic and semantic relationships in human language.
As a consequence, generative language models, such as ChatGPT, LLaMA~\cite{touvron2023llama}, and LAMDA~\cite{thoppilan2022lamda}, can produce machine-generated high-quality text that is indistinguishable from human writing. 
These generative capabilities have empowered these models to assist in diverse (creative) writing tasks and have been utilized to facilitate a wide range of interactive language-based applications within the HCI community~\cite{yuan2022wordcraft,petridis2023anglekindling,jiang2022promptmaker,wang2023enabling,wang2023popblends,osone2021buncho}.
For instance, WordCraft~\cite{yuan2022wordcraft} investigate the utilization of \acp{LLM} to aid fiction writers in tasks ranging from transforming a text to resemble a ``Dickensian'' style to providing suggestions to combat writer's block.
Their findings indicate that writers found such text-generating models beneficial even when the generated text is not perfect.
Petridis et al. \cite{petridis2023anglekindling} introduce AngleKindling, an interactive tool that employs \acp{LLM} to support journalists exploring different angles for reporting on a press release.
Their study with twelve professional journalists shows that participants found the system considerably more helpful and less mentally demanding than competitor brainstorming tools.
Other applications include prototyping support \cite{jiang2022promptmaker}, generation of titles and synopses from keywords \cite{osone2021buncho}, conversational interactions with mobile user interfaces \cite{wang2023enabling}, and conceptional blending \cite{wang2023popblends}.

Next to their text writing capabilities, language models are further known to retain commonsense knowledge within their training data, effectively transforming them into accessible knowledge stores that can be seamlessly queried using natural language prompts.
For instance, experiments with BERT~\cite{kenton2019bert} highlight that the model performance is competitive with traditional information extraction and open-domain question answering. 
Furthermore, recent studies show the potential of using ChatGPT for knowledge base construction, inspired by the fact that these language models have been pre-trained on vast internet-scale corpora that encompass diverse knowledge domains~\cite{wang2020language}.
However, it is worth noting that \acp{LLM} are known to frequently generate hallucinations, which are outputs that, while statistically plausible and seemingly believable, are factually incorrect~\cite{manakul2023selfcheckgpt,peng2023check}.

\subsection{Using LLMs to Support Programming}
The ability to generate new text and to reconstruct existing information makes \acp{LLM} highly appropriate to support users in software development, as programming often requires not only the creation of novel code segments tailored to current requirements and tasks but also depends on the application of established algorithms, software libraries, and best practices.
Accordingly, a large number of papers investigate \acp{LLM} specially trained for code generation~\cite{chen2021evaluating,clement2020pymt5,li2022competition,le2022coderl} as well as different approaches leveraging these models to provide interactive programmer support~\cite{jain2022jigsaw,jiang2022discovering,vaithilingam2022expectation,liu2023wants}.
For instance, Jiang et al. ~\cite{jiang2022discovering} discuss GenLine, a natural language code synthesis tool based on a generative \ac{LLM} and problem-specific prompts for creating or changing program code.
The findings from a user study indicate that the approach can provide valuable support to developers.
However, they also encounter several challenges, such as participants finding it difficult to form an accurate mental model of the kinds of requests that the model can reliably translate.
Similarly, Vaithilingam et al. \cite{vaithilingam2022expectation} conducted a user study with 24 participants evaluating their usage and experiences using the GitHub Copilot\footnote{\url{https://copilot.github.com} - \lastaccess} code generation model while programming.
The authors find that the synthesized code often provided a helpful starting point and saved online searching efforts. However, participants encountered issues with understanding, editing, and debugging code snippets from Copilot, resulting in not necessarily improved task completion times and success rates.
These findings align with the results of similar studies~\cite{dakhel2023github}.
However, a controlled experiment in \cite{kazemitabaar2023studying} records a positive effect of code generators when used in introductory programming courses for minors. 

The use of generative \acp{LLM} and code generators has been scarcely explored in scientific data analysis and not yet for scientific workflows.
Liu et al.~\cite{liu2023wants} examine the Codex code generator~\cite{chen2021evaluating} in the context of data analysis in spreadsheets for non-expert end-user programmers.
Moreover, several studies investigate the utilization for data visualization \cite{maddigan2023chat2vis,hassan2023chatgpt}.
For example, the study by Maddigan et al.~\cite{maddigan2023chat2vis} evaluates the efficiency of ChatGPT, Codex, and GPT-3 in producing scripts to create visualizations based on natural language queries.
The studies that have the most overlap with our work regarding the intention to support the design of data analysis pipelines are given by Ubani et al.~\cite{hassan2023chatgpt} and Zahra et al.~\cite{zahra2023laminar}.
In the case of the former, ChatGPT is used to build a conversational, natural language-based interface between users and the scikit-learn machine learning framework \cite{pedregosa2011scikit} supporting users in several phases of a machine learning project ranging from initial task formulation to comprehensive result interpretation.
For the latter, Laminar, a framework for serverless computing, is proposed, which offers possibilities for code searching, summarization, and completion.
However, the framework is solely focused on Python implementations.



\section{Methodology}

This section describes the study methodology. We begin by outlining about the general study approach. Then, we explain the research process for each experiment. Based on related work and the objectives of our research, we state the following research questions:

\begin{enumerate}
    \item[\textbf{RQ1:}] How performant is ChatGPT for comprehending and explaining scientific workflows?
    \item[\textbf{RQ2:}] How suitable is ChatGPT in suggesting and applying modifications for scientific workflows?
    \item[\textbf{RQ3:}] How efficient is ChatGPT in extending scientific workflows?
\end{enumerate}

\subsection{General Study Design}
To answer our research questions, we investigate the capabilities of ChatGPT, a widely-used LLM, to comprehend existing workflow descriptions (cf. Study I), to exchange tools used within a workflow (cf. Study II), and to extend a partially given workflow (cf. Study III) using three distinct user studies.
We select these use cases as understanding the data flow and the analysis performed is essential for successfully applying scientific workflows. 
Moreover, exchanging tools and extending a partially given workflow are common use cases in adapting and reusing existing workflows in the work context of domain scientists~\cite{cohen2011search}.  
For each study, we specially design conversational prompts simulating the interaction between a user working with workflows and ChatGPT.
For our studies, we leverage version GPT-3.5 of ChatGPT\footnote{\url{https://chat.openai.com} - \lastaccess}. We decided to use GPT-3.5 since it is openly available to the public and allows other researchers to reproduce our investigations without additional incurring costs\footnote{Moreover, we provide the chat records in the Supplementary Material of this article.}.
Additionally, we develop distinct questionnaires for evaluating the output of ChatGPT by the domain experts for each study.
While conducting a study, we present a brief overview of the study's overall goal and the developed questionnaire to the experts. 
Subsequently, the experts complete the questionnaire independently without the experimenters' support. This procedure is intended as participants were not pressured by a time limit and could freely allocate their time for the study. Furthermore, we intend to avoid a Hawthorne effect, where participants can provide biased responses due to the presence of observers~\cite{sedgwick2015understanding}.

\begin{table*}[!t]
    \centering
    \caption{Overview of the used workflows. We examine workflows from two scientific domains, i.e., bioinformatics and Earth observation, and two workflow systems (Nextflow and Apache Airflow). For each workflow, we report the number of (high-level) steps and used tools and in which study the workflow is used.}
        \begin{tabular}{clp{8.6cm}ccc}
            \toprule
            \textbf{Domain}&\textbf{Workflow}&\textbf{Description}&\textbf{\#Steps}&\textbf{\#Tools}&\textbf{Study}\\
                \hline

            \makecell[tl]{Bio-\\informatics}
    
                & \makecell[tl]{
                    WF1:\\crisprseq \cite{sanvicente2023crispr}\\
                    (\textit{Nextflow})
                }&
                \makecell*[t{{p{8.6cm}}}]{
                    A bioinformatics data pipeline for analyzing and evaluating gene editing experiments utilizing CRISPR-Cas9 mechanism for genome engineering. \\
                    Repository: \url{https://nf-co.re/crisprseq/2.0.0} (created 07/2022)
                }&
                6&
                9&
                I\\

            \cline{2-6}
            
            &
                \makecell[tl]{
                    WF2:\\RS-Star\\
                    (\textit{Nextflow})
                }&
                \makecell*[t{{p{8.6cm}}}]{
                    The general aim of this workflow is to perform differential gene expression analysis using RNA-seq data.\\
                    Repository: \url{https://github.com/Nine-s/nextflow_RS1_star} (created 11/2021)
                }&
                5&
                5&
                I,II,III\\

            \hline 

            \makecell[tl]{Earth\\Observation}&
                \makecell[tl]{
                    WF3:\\FORCE2NXF-\\Rangeland \cite{lehmann2021force}\\
                    (\textit{Nextflow})
                }&
                \makecell*[t{{p{8.6cm}}}]{
                    This workflow analyzes long-term vegetation dynamics in the Mediterranean using the FORCE-toolkit. \\
                    Repository: \url{https://github.com/CRC-FONDA/FORCE2NXF-Rangeland} (created 11/2020)
                }&
                9&
                8&
                I\\

            \cline{2-6}

            &
                \makecell[tl]{
                    WF4:\\Grasslands\\
                    (\textit{Nextflow})
                }&
                \makecell*[t{{p{8.6cm}}}]{
                    This workflow aims at understanding differences in long-term changes (1984-2022) in ground cover fractions specific to European grasslands. \\
                    Repository: \url{https://github.com/kelewinska/FONDA_trends-nf} (created 08/2023)
                }&
                6&
                3&
                I,III\\
                
            \cline{2-6}
            &
                \makecell[tl]{
                    WF5:\\FORCE\\
                    (\textit{Apache Airflow})
                }&
                \makecell*[t{{p{8.6cm}}}]{
                    This workflow focuses on analyzing long-term vegetation dynamics in the Mediterranean using the FORCE-toolkit.\\
                    Repository: \url{https://github.com/CRC-FONDA/fonda-airflow-dags/tree/main} (created 02/2021)
                }&
                8&
                8&
                I\\
                \bottomrule
        \end{tabular}
    \label{tab:workflows}
\end{table*}

\subsection{Participants}
Throughout all experiments, we recruited one expert from bioinformatics and three experts working on Earth observation workflows. Using scientific workflows is common in these two areas. In bioinformatics, scientific workflows are an important tool for enabling the automation and documentation of complex data analysis processes, ensuring reproducibility and transparency in research~\cite{di2017nextflow}. 
In Earth observation, scientific workflows streamline the complex process of acquiring, processing, and analyzing vast amounts of satellite and sensor data, enhancing the efficiency and accuracy of environmental studies~\cite{sudmanns2020big}. Hence, scientific workflows have become commonplace in these two areas.
%
The professions include postdocs and PhD students working at universities. All participants hold a master's degree in their profession and several years of experience in their domain.
%
%
%
%
All experts are between 25 and 40 years old (two female, two male).
\subsection{Scientific Workflows}
In our study, we consider a total of five different workflows. We summarize the used workflows and their details in Table~\ref{tab:workflows}.
The workflows are taken from the work context of the recruited experts, given their high degree of familiarity and expertise with them.
In bioinformatics, we use two workflows that deal with the analysis of genomic data.
First, the \textit{crisprseq} workflow, sourced from the \textit{nf-core} repository\footnote{\url{https://nf-co.re} - \lastaccess}, a hub for best-practice workflows, focuses on analyzing and evaluating gene editing experiments utilizing CRISPR-Cas9 mechanism for genome engineering.
Second, the \textit{RS-STAR} workflow, which was implemented by the recruited domain expert and performs differential gene expression analysis using RNA-seq data. 
The \textit{FORCE2NXF-Rangeland} and \textit{FORCE} are two implementations of an Earth observation workflow that is concerned with analyzing long-term vegetation dynamics in the Mediterranean using the FORCE toolkit\footnote{\url{https://github.com/davidfrantz/force} - \lastaccess}, which provides processing routines for satellite image archives.
The former is implemented using the Nextflow scientific workflow management system\footnote{\url{https://www.nextflow.io} - \lastaccess}~\cite{di2017nextflow} and the latter by leveraging Apache Airflow\footnote{\url{https://airflow.apache.org} - \lastaccess}~\cite{harenslak2021data}.
The third Earth observation workflow called \textit{Grasslands}. builds on previous work~\cite{lewinska2021changes,lewinska2020short} aiming at understanding differences in long-term changes in-ground cover fractions specific to European grasslands depending on the definition of endmembers (i.e., unique spectral signatures of a specific material or ground cover) approximating these fractions.

\subsection{\ac{LLM} Prompting}
The choice and design of prompts entered into a \ac{LLM} has a decisive influence on the output quality~\cite{white2023prompt} and, in our case, on the suitability of ChatGPT for workflow development and implementation. 
Our prompts are organized first to provide the context, often including the workflow script, followed by the specific question or instruction under investigation.
Suppose the workflow is divided into sub-workflows, possibly distributed over several files. In that case, we first specify the main workflow and then all sub-workflows and task definitions in the order they occur in the main workflow. 
In our research's initial stages, we experimented with various alternative prompts for each user study, incrementally modifying and enhancing them in response to the outcomes we received.
For example, for Study~I (i.e., workflow comprehension), we discovered that ChatGPT tends to describe properties of the workflow language or technical aspects instead of workflow characteristics.   
Such phenomena could be resolved by adding explicit instructions, e.g., ``\textit{do not explain nextflow concepts}''.
We refer to Section~\ref{sec:prompt-design-challenges} for a detailed discussion of prompt design challenges.
We stopped this adjustment process after a few iterations as soon as no more of such artificial artifacts were generated.
We have refrained from more extensive prompt engineering since workflow designers are domain experts from diverse fields who cannot be assumed to be specialists in developing and tuning prompts.
However, we acknowledge that the choice of wording in our prompts influences the results~\cite{white2023prompt}. 
%
%
We discuss the limitations of our work concerning the chosen study design and prompt strategy selection in more detail in Section~\ref{sec:limitations}.

\section{Study I: Workflow Comprehension}
\label{sec:study_i}

\begin{table*}[tbhp]
    \centering
    \caption{Overview of the used prompts to investigate ChatGPT's capabilities in capturing the content of a workflow description (Study I). \textit{[workflow-language]} and \textit{[workflow-text]} represent placeholders for the workflow management system, i.e., Nextflow or Apache Airflow, and the workflow description text. All prompts are executed within one conversation.
    } 
    \begin{tabular}{cp{2.0cm}p{14cm}}
            \toprule
        \textbf{ID}&\textbf{Category}&\textbf{Prompt}\\
            \midrule
            P1\_1&
            \makecell[{{p{2.0cm}}}]{\centering Overall\\aim}&
            \makecell*[{{p{14cm}}}]{%
                The following text contains a scientific workflow written in \textit{[workflow-language]}: \\[0.2em]
                \textit{[workflow-text]} \\[0.2em]
                Explain from which research area this workflow originates and describe the general aim of this workflow. Don't explain \textit{[workflow-language]} concepts.
            }\\
        \hline
            P1\_2&
            \makecell[{{p{2.0cm}}}]{\centering Workflow\\explanation}&
            \makecell*[{{p{14cm}}}]{%
                Explain all individual tasks that are implemented in this workflow. For each task explain which software programs or tools are used in this workflow to perform the task. Don't explain \textit{[workflow-language]} concepts.
            }\\
        \hline
            P1\_3&
            \makecell[{{p{2.0cm}}}]{\centering Workflow\\explanation}&
            \makecell*[{{p{14cm}}}]{%
                Explain the type of input data and the format of the input data needed for this workflow. Don't explain the workflow itself.
            }\\
        \hline
            P1\_4&
            \makecell[{{p{2.0cm}}}]{\centering Workflow\\explanation}&
            \makecell*[{{p{14cm}}}]{%
                Explain the overall result of this workflow. For each individual task of the workflow report the type of data that is produced by this task. 
            }\\
        \hline
            P1\_5&
            \makecell[{{p{2.0cm}}}]{\centering Research\\questions}&
            \makecell*[{{p{14cm}}}]{%
                Explain up to 3 research questions for which this workflow is helpful.
            }\\
            \bottomrule
    \end{tabular}
    \label{tab:prompts_us1}
\end{table*}

In our first study, we investigate the capabilities of ChatGPT in capturing the actual purpose of a workflow. In other words, we are prompting ChatGPT to explain the purpose of a workflow. In this study, we assess ChatGPT's quality in comprehending and explaining a workflow's purpose in a user study involving workflow experts.
Understanding the data flow and the analysis performed constitutes an important aspect of the daily work with scientific workflows.
On the one hand, workflows are often precisely adapted to individual research questions, which makes it challenging even for other experts from the same domain to understand them.
On the other hand, in many research institutions, an increasing number of legacy workflows whose original authors and contributors are no longer available for maintaining and refining the codes require taking over by new team members. 
Understanding a workflow is usually a prerequisite for adopting and applying a workflow correctly.
%
%

Thus, Study~I pursues three goals: how well does ChatGPT perform on (a) identifying the domain and the overall objective of the analysis, (b) reporting the individual computation steps, used tools, their needed input data, produced output data, and (c) explaining research questions for which these analyses are helpful given the workflow description.
The first two parts of the study have a reconstructive character, whereas the third is more explorative, requiring ChatGPT to reason beyond the given workflow description.
We build a set of five different prompts to evaluate ChatGPT's capabilities concerning the three dimensions. 
When providing the workflow definition in the prompt, delete all comments within the definition to prevent information leakage.
Table \ref{tab:prompts_us1} depicts the developed prompts.
\begin{figure*}[!tbhp]
    \centering
    \subfloat[][]{
        \includegraphics[width=0.65\textwidth]{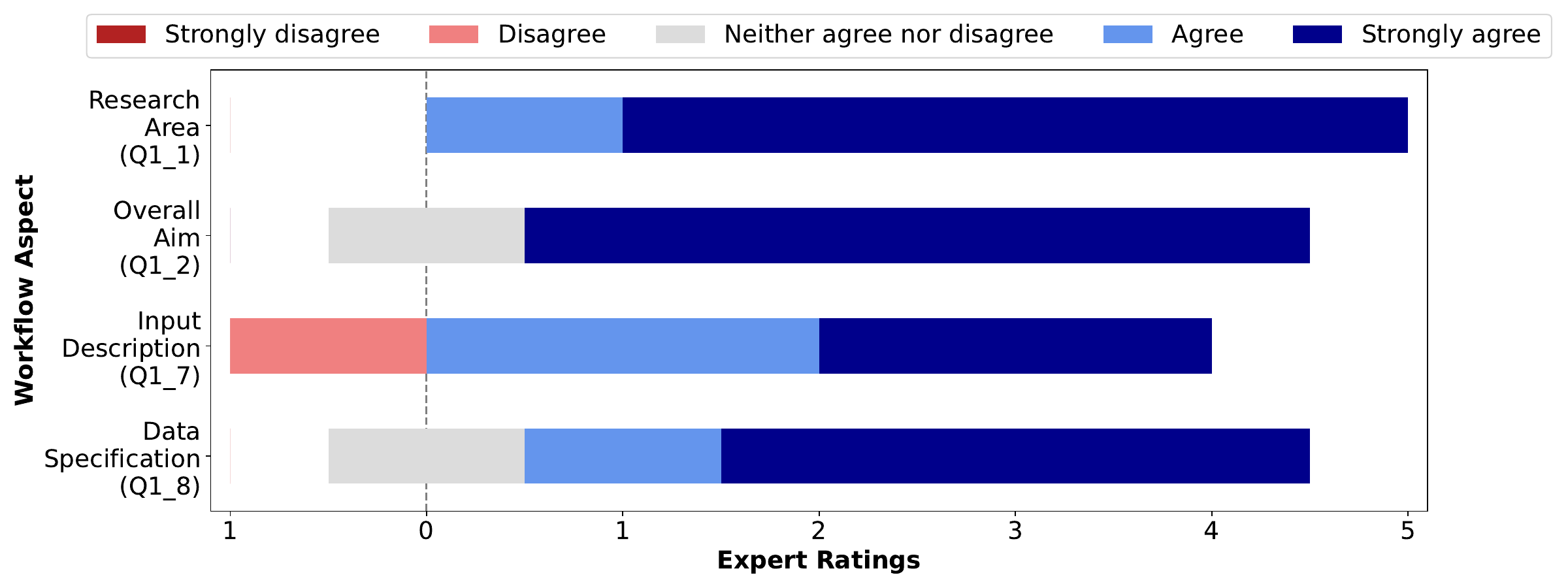}
        \label{fig:us1_likert}}
    \subfloat[][]{
        \includegraphics[width=0.32\textwidth]{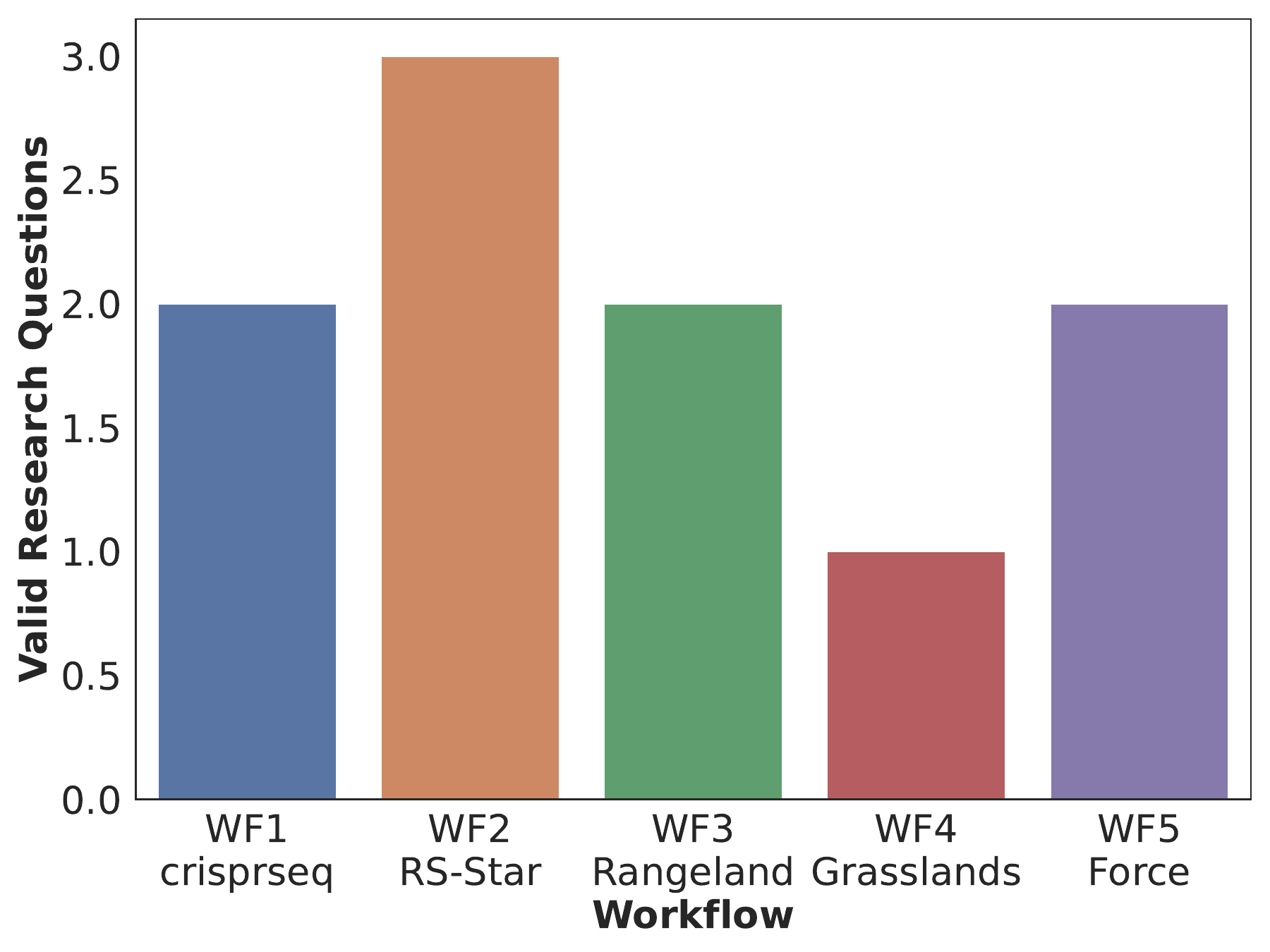}
        \label{fig:us1_rq}}
    \caption{Results from Study I: (a) rating distribution of the domain experts for ChatGPT's capability in identifying the research area and overall aim as well as the input and data description of a workflow. The question item identifier (see \ref{sec:appendix-form-us1}) is given in parenthesis for each row. (b) Number of valid research questions generated by ChatGPT for the different workflows as assessed by the domain experts (question item \textit{Q1\_9}). We prompted ChatGPT to output up to three research questions per workflow.}
    \label{fig:us1_likert_us1_rq}
\end{figure*}
For each workflow, all prompts are executed in one conversation, enabling ChatGPT to use the in- and output of previous prompts as context information.
We ask the domain scientists to evaluate answers given by ChatGPT using a feedback questionnaire.
The complete questionnaire contains nine items in total and can be found in full-length in ~\ref{sec:appendix-form-us1}. The questionnaire focuses on the correctness of the prompts regarding the aim of the workflow, the explanation, and the forecast of addressed research questions.
For four of the nine items, users rate the generated explanations on a 5-point Likert scale\footnote{1: Strongly Disagree; 5 Strongly Agree.}.
In addition, three items comprise quantitative evaluations of how many computational steps are correctly detected, how many utilized software tools and programs are accurately identified, and how many valid follow-up research questions.
The remaining question items concern the quality of the explanations of the workflow sequence, the description of the tools used, and the results produced. 
We add a comment field for each item to report issues and errors in the generated explanations if the domain expert does not apply the content. 

\subsection{Results}
The results of the expert surveys are presented in the following according to the three subcategories of the questionnaire, i.e., research area and the overall aim of the workflow, explanation of workflow details, and subsequent research questions.
\subsubsection{Overall Aim of the Workflow}
The first two rows of Figure~\ref{fig:us1_likert} highlight the rating distribution of the domain experts concerning ChatGPT's explanations for the research area and the overall aim of the workflows.
The experts agreed with the explanations generated, indicating a basic understanding of ChatGPT regarding the different workflows ($\mu=4.7$, $\sigma=0.68$).
The lowest agreement, with a score of 4 concerning the research area and 3 for the overall aim, was recorded in the evaluation of the \textit{WF4-Grasslands} workflow. 
In this case, the expert could not agree with the explanations mainly due to the wrong interpretation of an abbreviation within the workflow description, i.e., \textit{FNF} was misinterpreted as \textit{``fraction of non-forest''}, instead of \textit{fold and fill}.
This misunderstanding resulted in the workflow being explained as examining forest regions rather than grasslands.

\subsubsection{Workflow Explanation}
%

\begin{figure*}[!tbhp]
    \centering
    \subfloat[][]{
        \includegraphics[width=0.49\textwidth]{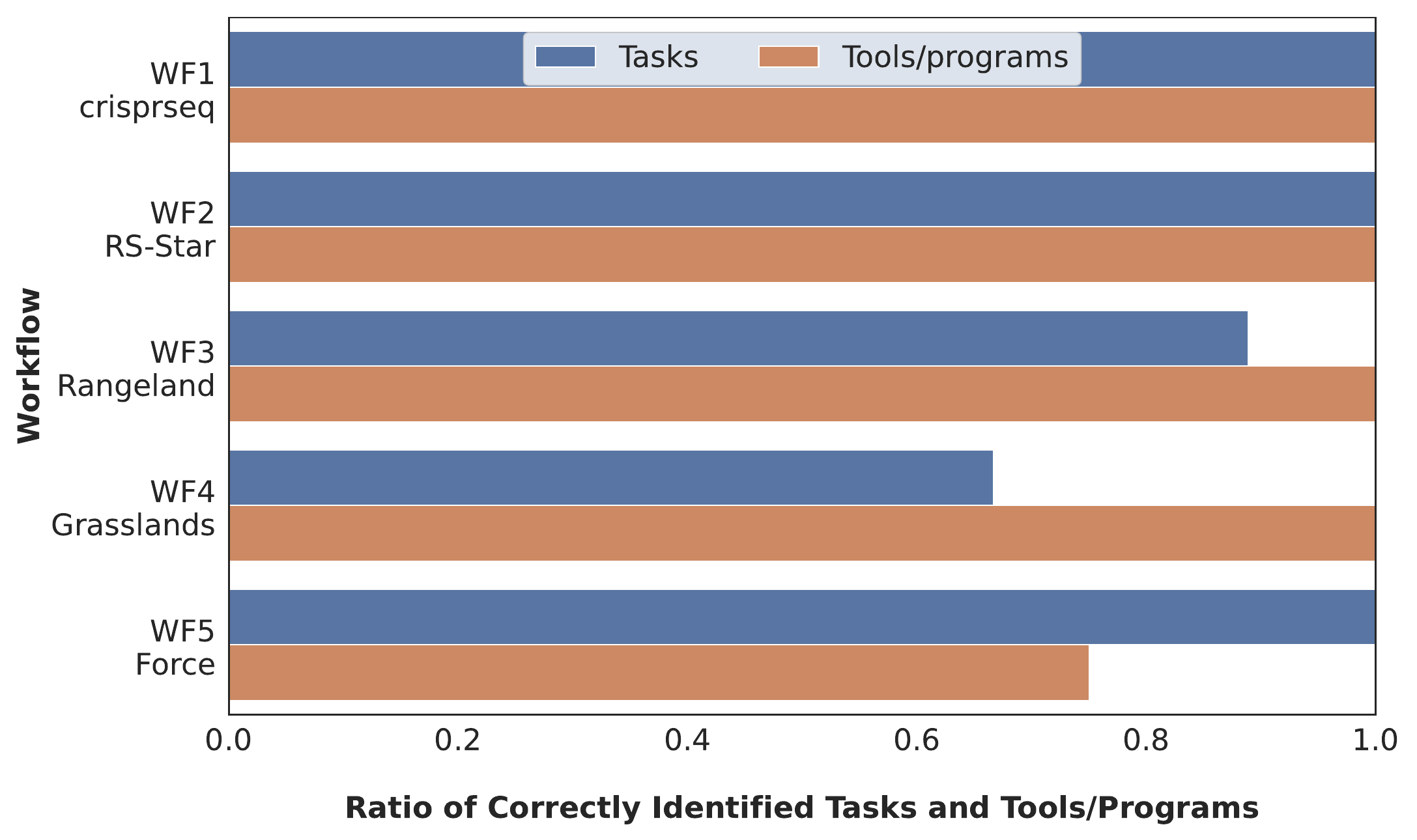}
        \label{fig:us1_tt_quant}}
    \subfloat[][]{
        \includegraphics[width=0.49\textwidth]{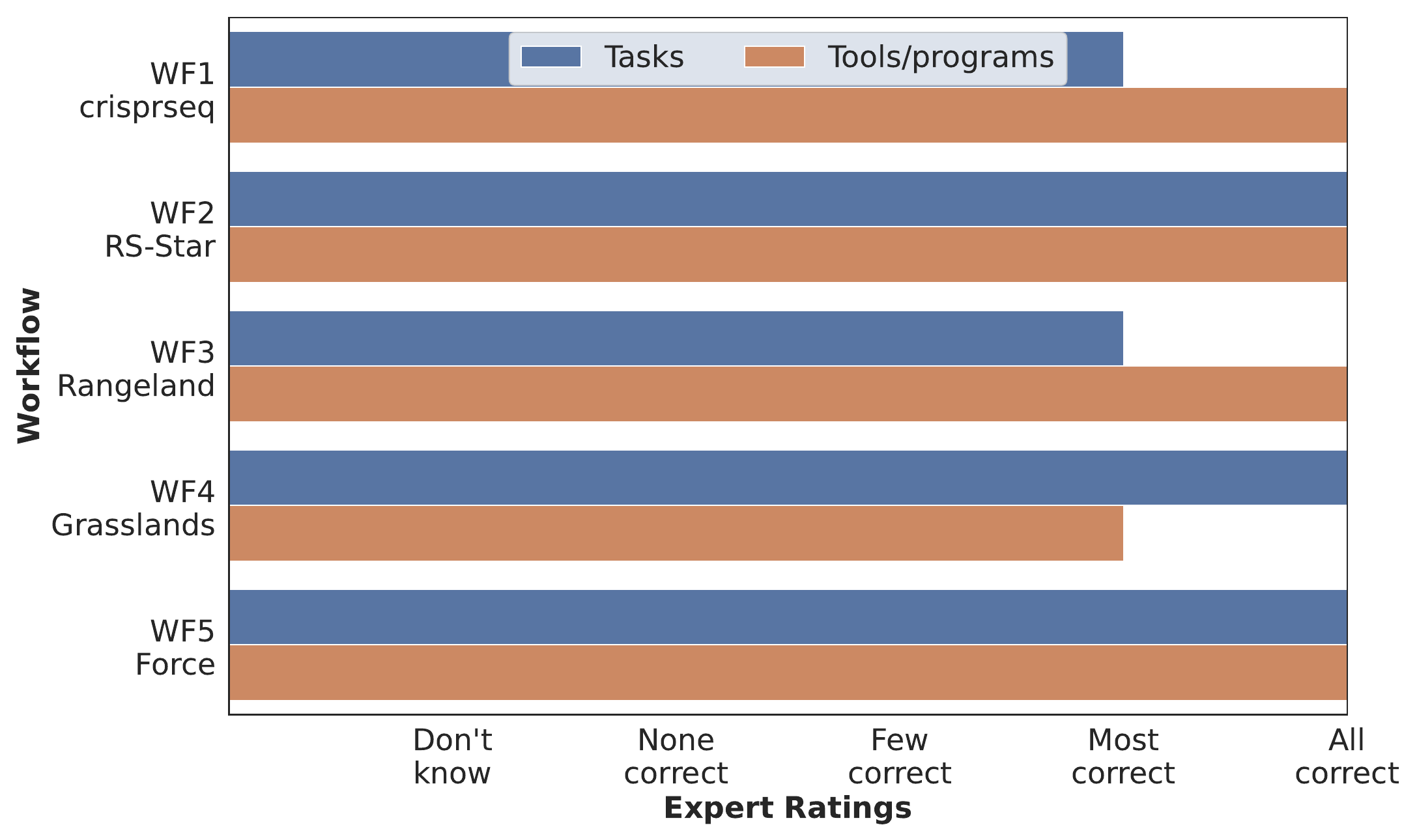}
        \label{fig:us1_tt_expl}}
    \caption{Result statistics highlighting the number of correctly identified tasks and tools of a workflow (a) and their explanation (b). We report separate results per investigated workflow. The results are based on the ratings of the domain experts. Results in (a) correspond to the items \textit{Q1\_3} and \textit{Q1\_5} and in (b) to \textit{Q1\_4} and \textit{Q1\_6} of the questionnaire given in \ref{sec:appendix-form-us1}.}
    \label{fig:us1_tt_quant_us1_tt_expl}
\end{figure*}
In general, the participants regard the quality of the explanations given by ChatGPT as high (see Figures~\ref{fig:us1_tt_quant} and \ref{fig:us1_tt_expl}).
All computational steps are accurately identified in three of the five workflow descriptions. 
Moreover, for four of these five workflows, every tool employed is correctly detected, but in \textit{WF5-Force}, two out of eight were missing.
The detailed descriptions of the tasks and tools provided by ChatGPT were also judged to be coherent by the experts. 
Overall, the worst performance is achieved with the output for workflow \textit{WF4-Grasslands} for which only four out of six tasks are correctly extracted and only most of the tools are correctly described. 
These errors are mainly due to follow-up errors that result from the incorrect recognition of the workflow purpose. 

The results of the questions items (see Q1\_7 and Q1\_8 in ~\ref{sec:appendix-form-us1}), which assess the produced information about the format and type of input and output data of ChatGPT, can be seen in the two lower rows in Figure~\ref{fig:us1_likert}.
Similar to the previous findings, the information generated for four of the five workflows are evaluated positively ($\mu=4.2$, $\sigma=1.0$).
Again, only the produced information for the workflow \textit{WF4} was assessed as neutral or negative, i.e., input description score of 2 (disagree) and data specification of 3 (neither agree nor disagree).
\subsubsection{Research Questions}
The last query was concerned with explaining up to three subsequent research questions to a given workflow.
Figure~\ref{fig:us1_rq} shows the result of this question item.
ChatGPT achieves only moderate performance, generating just for one workflow, i.e., \textit{WF2-RS-Star}, three valid research questions.
In total, out of the 15 generated research questions 10 were correct.
These figures suggest that ChatGPT offers only a reduced performance for more explorative tasks.

\section{Study II: Workflow Modification}
\label{sec:study_ii}
%
The second study investigates how much ChatGPT can aid domain experts in modifying and tailoring a workflow in our second study.
Researchers usually do not start from scratch when developing workflows but typically adapt or reuse parts of existing workflows from the community~\cite{cohen2011search}.
This strategy applies in particular to biomedicine, in which workflows are more widespread and have a longer tradition compared to other domains~\cite{oinn2004taverna}.
For example, genomic workflows will often be applied to a broad spectrum of data originating from different sources, each with its distinctive features and characteristics, making it necessary to adjust the workflow definition for more efficient data processing.
Moreover, technological advancements, such as in genome sequencing technology~\cite{van2014ten}, lead to new tools specially developed to leverage the capabilities of the new technologies.
The continuous integration of new and alternative scientific tools into existing workflows is essential to conduct state-of-the-art research~\cite{hu2021next}.

In our study, we are particularly investigating the exchange of used tools in the bioinformatics workflow \textit{WF2-RS-Star} whose computational scheme is given in Figure~\ref{fig:ninon_a}.
We select two parts of the workflow to be modified:
\begin{enumerate}
    \item read trimming and filtering (also called read quality control), originally performed by FASTP~\cite{chen2018fastp} and
    \item reference genome indexing and alignment, carried out by STAR~\cite{dobin2013star}. 
\end{enumerate}
For assessing the workflow modification capabilities of ChatGPT, we build four prompts (see Table~\ref{tab:prompts_us2}).
The first prompt requests a list of alternative tools for a given workflow step from ChatGPT.
The second and third prompts request the recommendation of two alternative tools, including an explanation of the suggestion, a comparison of the selected tools with the tool originally used in the workflow script, and their strengths and weaknesses.
With the last prompt, the actual rewriting of the workflow to include the selected tool is requested.
We test the inclusion of two alternative tools per computational task, i.e., the prompts \textit{P2\_3} and \textit{P2\_4} (see Table \ref{tab:prompts_us2}) are carried out once for each tool from the recommendation.
Analogous to Study I, we use a questionnaire for evaluating ChatGPT's output by the biomedical domain expert containing 13 items in total.
For most items (9 out of 13), the generated texts and explanations concerning methodical differences or pros and cons of the tools should be rated on a 5-point Likert scale. 
The remaining questions require numerical ratings~(2 items), yes/no answers~(1), and free text fields~(1).
Again, we add a comment field for each item to report issues and errors in the generated explanations if the domain expert does not fully apply the content. 
The complete questionnaire is given in ~\ref{sec:appendix-form-us2}.
When conducting the study, we also provide the generated workflow scripts to the domain experts and asked them to execute them on their systems.
Furthermore, we request the experts to inspect and correct any non-functional scripts. 
For the latter, we set a time limit of 20 minutes per tool substitution.

\subsection{Results}
The results are presented in the following according to the two subcategories of the prompts, including the exploration of alternative tools and workflow modification. 
We summarize the results of the two use cases (i.e., read quality control, reference genome alignment) and each of the two alternative tools when reporting the results.
\subsubsection{Tool Exploration}
\begin{figure*}[tbhp]
\includegraphics[width=0.95\textwidth]{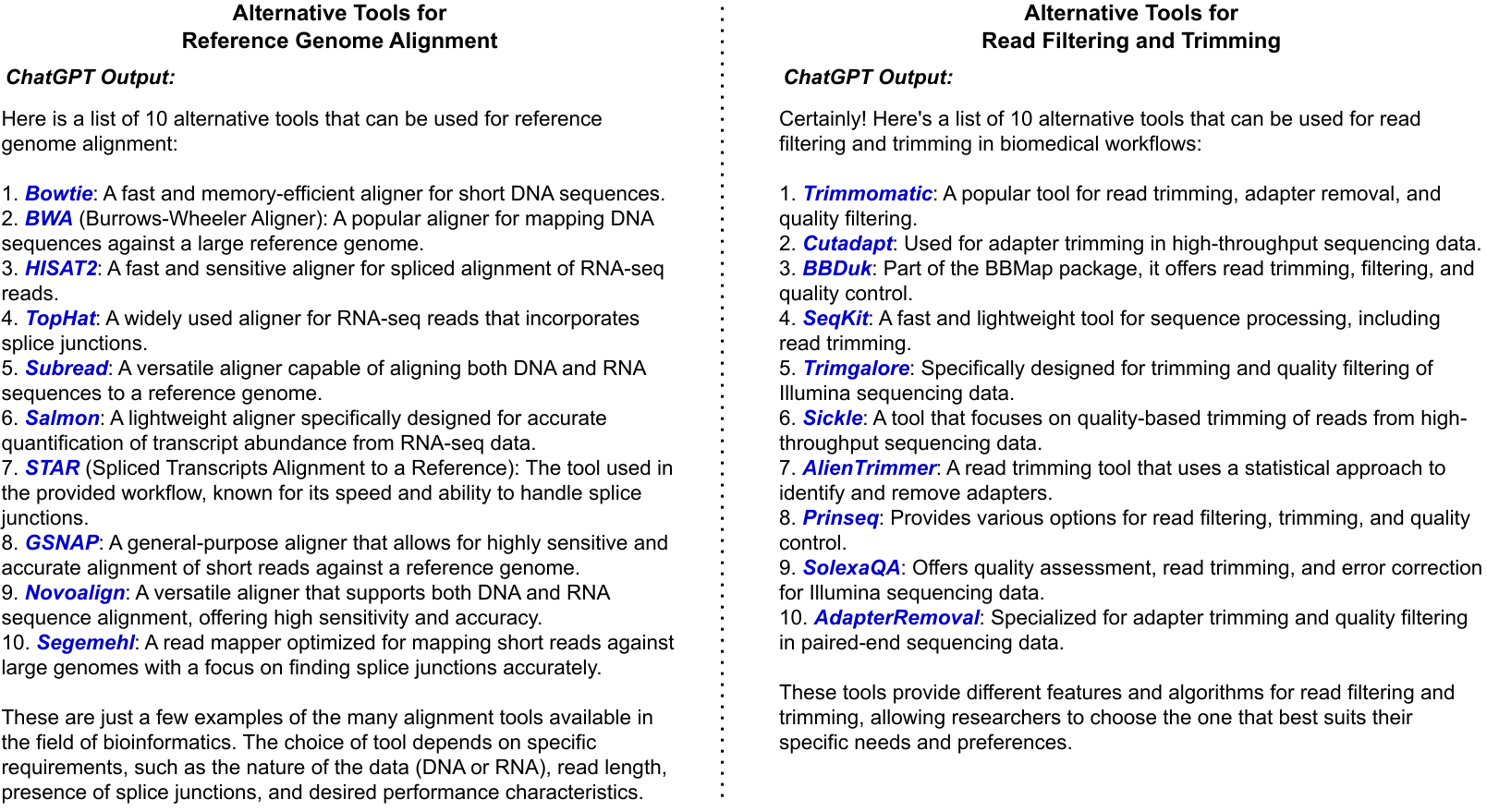}
\centering
\caption{Representation of the output of ChatGPT when requested to provide a list of alternative tools for reference genome indexing and alignment (left) and read quality control (right) using prompt P2\_1. All tools are assessed to be valid by the biomedical domain expert.}
\label{fig:us2_alt_tool_lists}
\end{figure*}

\begin{figure*}[tbhp]
\includegraphics[width=0.90\textwidth]{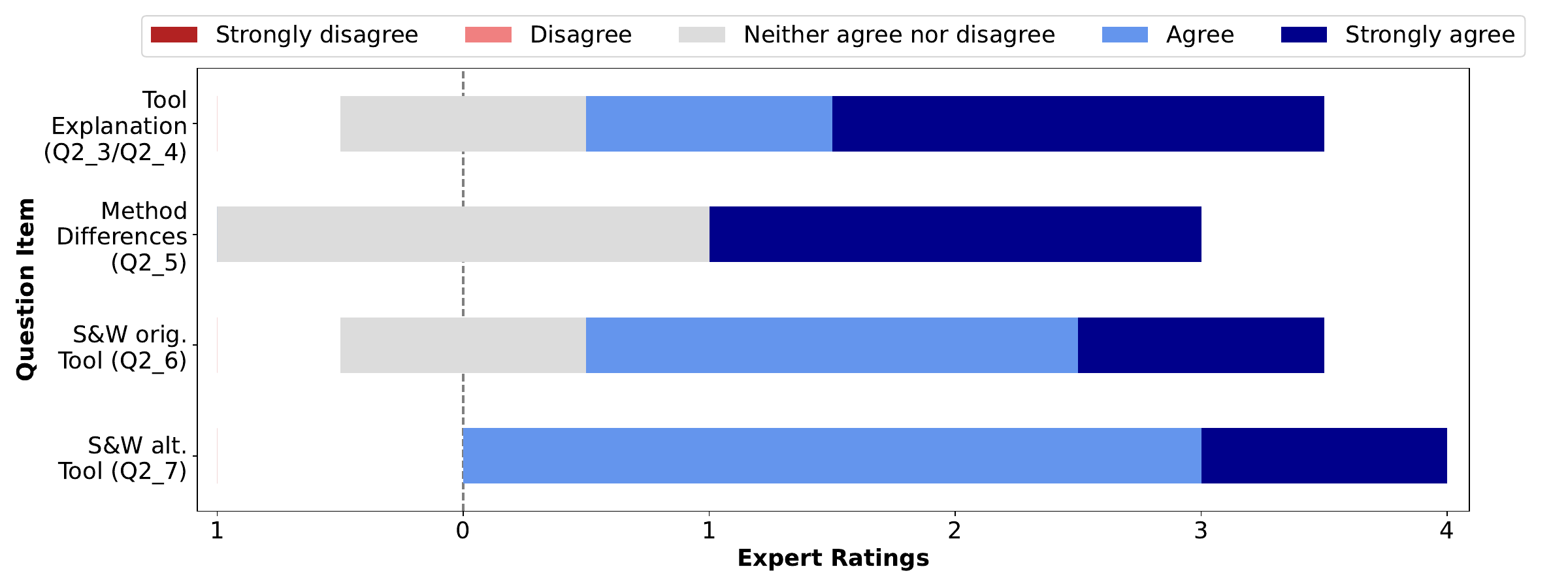}
\centering
\caption{Overview of the rating distribution of the biomedical domain expert for ChatGPT's capability for explaining alternative tools, methodical differences, and strengths and weaknesses (S\&W) of the tools. The question item identifier (see \ref{sec:appendix-form-us2}) is given in parenthesis for each row.}
\label{fig:us2_likert}
\end{figure*}

When exploring possible alternative tools, ChatGPT showed a good performance, providing a fully valid list of ten alternative tools using prompt \textit{P2\_1} (see Table~\ref{tab:prompts_us2}) for both scenarios.
The generated output list for both tasks can be seen in Figure~\ref{fig:us2_alt_tool_lists}.
However, the domain expert criticized the alternative tools for reference genome alignment specified Bowtie~\cite{langmead2009ultrafast} and not its successor Bowtie2~\cite{langmead2012fast}.
\begin{table*}[t]
    \centering
    \caption{Overview of the used prompts to investigate ChatGPT's capabilities in swapping used tools in bioinformatic workflows (Study II). 
    Information in square brackets specifies placeholders for concrete information regarding the workflow or the tool to be replaced.
    } 
    \begin{tabular}{cp{2.1cm}p{14.0cm}}
            \toprule
        \textbf{ID}&\textbf{Category}&\textbf{Prompt}\\
            \midrule
            
            P2\_1&
            \makecell[{{p{2.1cm}}}]{\centering Tool\\exploration}&
            \makecell*[{{p{14.0cm}}}]{%
            The following text contains a \textit{[domain]} workflow written in \textit{[workflow-language]}: \\[0.2em]
            \textit{[main-workflow]} \\[0.2em]
            The following snippets contain the source code for the step of the workflow which uses \textit{[tool]} to perform \textit{[step]}: \textit{[step-source-code]}. Please provide a list of 10 alternative tools to perform \textit{[tool]}.  
            } \\
        \hline

            P2\_2&
            \makecell[{{p{2.1cm}}}]{\centering Tool\\exploration}&
            \makecell*[{{p{14.0cm}}}]{%
            The following text contains a \textit{[domain]} workflow written in \textit{[workflow-language]}: \\[0.2em]
            \textit{[main-workflow]} \\[0.2em]
            The following snippets contain the source code for the step of the workflow which uses \textit{[tool]} to perform \textit{[step]}: \textit{[step-source-code]}. Alternative tools for \textit{[step]} are: \textit{[list-of-tools]} \\
            Which of the tools would you recommend as most suitable alternative for \textit{[step]} in the given workflow. Please name the two alternatives and give an explanation why these tools are especially advisable for the given workflow.  
            } \\
        \hline
        
            P2\_3&
            \makecell[{{p{2.1cm}}}]{\centering Tool\\exploration}&
            \makecell[{{p{14.0cm}}}]{%
            \textit{[original-tool]} and \textit{[alternative-tool]} are two tools for \textit{[step]} in \textit{[domain]} workflows. First,  explain the differences between the tools and the underlying approaches. Second, name strenghts and weaknesses of both tools.
            } \\

        \hline

            P2\_4&
            \makecell[{{p{2.1cm}}}]{\centering Workflow\\modification}&
            \makecell*[{{p{14.0cm}}}]{%
            The following text contains a \textit{[domain]} workflow written in \textit{[workflow-language]}: \\[0.2em]
            \textit{[main-workflow]} \\[0.2em]
            The following snippets contain the source code for the step of the workflow which uses the \textit{[tool]} to perform \textit{[step]}: \textit{[step-source-code]} \\
            Please re-write the code of the workflow and the proccess to use \textit{[alternative-tool]} instead of \textit{[original-tool]}. The number of parameters of the individual process descriptions may have to be adjusted. Please explain features / options of \textit{[original-tool]} which are not supported in \textit{[new-tool]}.  
            } \\

            \bottomrule
    \end{tabular}
    \label{tab:prompts_us2}
\end{table*}
From these, ChatGPT recommended HISAT2~\cite{kim2019graph} and Bowtie as alternatives for genome alignment and Trimmomatic~\cite{bolger2014trimmomatic} and Cutadapt~\cite{martin2011cutadapt} for read quality control.
According to the domain expert, all tools are principally valid substitutions, but Bowtie is a rather inappropriate suggestion since it is not specialized for RNAseq data. 
The results of explaining the tools' methodical differences and strengths and weaknesses can be seen in Figure~\ref{fig:us2_likert}.
Essentially, the generated textual explanations were assessed positively, except when describing the methodological differences between the tools.
In this case,  ChatGPT could offer a convincing explanation in only 50\% of the cases.
In the other cases, the texts were too general, and differences were not named clearly.
In summary, similar to the results from Study I, ChatGPT solves these rather reconstructive tasks well, reaching an average score of 4.1 (min=3, max=5, $\sigma$=0.81) when taking the results from items Q2\_3 to Q2\_7 (see questionnaire in \ref{sec:appendix-form-us2}) into account.
These results indicate that ChatGPT could effectively used to explore tools in a given field.

\subsubsection{Workflow Modification}
We requested ChatGPT to re-write the workflow script for each of the two recommended alternative tools in both use cases, resulting in four modified workflows in total. 
Table~\ref{tab:us2_results} summarizes the results achieved. 
First of all, none of the generated workflow scripts was entirely correct.
Only in one of the four cases, the generated script was (at least) syntactically valid, i.e., if replacing FASTP with Trimmomatic for read quality control, and the domain expert could execute it without further adaptations.
However, the script was not a semantically correct modification of the original workflow script since one particular quality control routine (i.e., PolyG trimming) was not reflected in the adopted script.
This issue also occurred for Cutadapt, the other alternative tool for quality control.
For both tools, ChatGPT reported in the explanation text that the tools do not support this feature; however, in reality, they do.
This failure could be interpreted as a kind of LLM hallucination \cite{manakul2023selfcheckgpt}.
The second task, reference genome indexing and alignment, revealed different issues than the first.
Here, the main problem was the correct linking of the two sub-parts of the task, first the index generation and then the computation of the actual alignment.
For the former, each tool specifies and uses its distinct data format and defines how to store the index (e.g., saving it in one or multiple files).
However, the storing strategy also affects how the output of the indexing task has to be passed on to the input of the alignment computation. 
In the scripts generated by ChatGPT, the actual step descriptions to invoke indexing and alignment by the tools were (generally) valid. However, the linking of these two needed to be corrected.
For example, Bowtie saves its index in multiple files sharing a common file name prefix, which has to be specified as a parameter during alignment.
However, in the modified script, the list of all files of a specially created directory was passed to the alignment process.
For Bowtie, this problem could be easily fixed by the domain expert, but for HISAT2, it was not that trivial and hence could not be solved in the given time budget of 20 minutes. 

To sum up, the study's results indicate that modifying workflow scripts poses considerable challenges for ChatGPT as it requires a detailed understanding of the tool's idiosyncrasy, the exact computations they perform, and the data formats they use.

\section{Study III: Workflow Extension}
\label{sec:study_iii}
In the third study, we investigate the capabilities of ChatGPT in extending a scientific workflow given a partial script. 
As discussed in the motivation for Study~II (see Section~\ref{sec:study_ii}), users often reuse parts of existing workflows from the research community and adapt them to the research question at hand by enhancing the pipeline with additional analyses and computational steps~\cite{cohen2011search}.
Moreover, data analysis projects are often exploratory processes, and computation pipelines are incrementally adapted and extended based on executions and findings from previous versions of the workflow, e.g., to include additional data correctness checks, add more differentiated result evaluations and provide advanced result visualizations~\cite{zeyen2019adaptation}.  
In our study, we simulate this incremental exploration process by taking an existing workflow and removing \textit{n} steps at the end of it. 
We then request ChatGPT to a) enumerate the necessary steps to accomplish the original research goal and b) regenerate the next step using the tool of the original pipeline or by giving a verbal description of the task.
For this study, we select one workflow from each research domain for investigation: \textit{WF2-RS-Star} for biomedicine and \textit{WF4-Grasslands} for earth observation.
The two workflows were chosen because they offer different implementation characteristics, i.e., \textit{WF2-RS-Star} leverages almost exclusively external tools, whereas \textit{Grasslands} relies more strongly on specially implemented R and Python scripts.
Moreover, Study~I (see Section~\ref{sec:study_i}) already showed notable result differences of ChatGPT for both workflows.
By choosing these two specific workflows, we aim to encompass a possibly broad spectrum of performance variations.
We test ChatGPT's workflow extension capabilities in three scenarios:
For WF2-RS-Star, we remove the last step \textit{transcript quantification} as well as the last two steps \textit{transcript quantification} and \textit{format conversion}, forming two extension scenarios.
In the case of \textit{WF4-Grasslands}, we remove all steps at the tail of the workflow, including \textit{autoregressive trend analysis} (see schema in \ref{sec:appendix-grasslands}).

Table~\ref{tab:prompts_us3} illustrates the prompts developed for this purpose.
This study uses slightly different prompts (see P3\_2a and P3\_2b) reflecting the different workflow types, i.e., tool- vs. script-based.
For the latter, we include additional instructions to a) specify the programming language of the script and b) ask the domain expert for a verbal description of the computational steps to be implemented.
See \ref{sec:appendix-verbal-description} for the verbal description provided by the earth observation expert.
The questionnaire for evaluating the generated outputs consisting of seven items can be found in \ref{sec:appendix-form-us3}.
\subsection{Results}
The results are presented in the following according to the two subcategories of the prompts, i.e., workflow exploration and extension. 

\subsubsection{Workflow exploration}
For describing further computational steps necessary to accomplish a specific research goal given a partial workflow, ChatGPT showed mixed results.
The LLM provides a correct list of suitable steps in two of the three scenarios. 
Also, the tools and methods for implementing the steps suggested by ChatGPT were valid.
However, both domain experts criticize that the specifications for the necessary steps and the proposed tools tend to be rather generic and generalized.
For instance, for extending the \textit{WF4-Grasslands} workflow the earth observation expert commented:
\begin{quote}
    \textit{Overall, the proposed workflow is very generic and does not provide a clear roadmap for the analyses. It also proposes to use very simplistic and often imperfect approaches.}
\end{quote}
Overall, the results confirm the findings from the two previous studies that ChatGPT shows weaknesses in more exploratory tasks.
\subsubsection{Workflow extension}
Using the prompts P3\_2a and P3\_2b (see Table~\ref{tab:prompts_us3}), we request ChatGPT to re-construct the last removed computational step in each extension scenario.
Table~\ref{tab:us3_results} summarizes the results achieved.
Like the results from Study II, ChatGPT shows considerable weaknesses in the automatic extension of workflows.
None of the generated workflow scripts was executable without the intervention of the domain expert.
A clear difference is revealed when comparing the two domains, biomedicine and earth observation.
In the former case, the generated workflow scripts are (at least) of such a quality that the domain expert could successfully correct them within 20 minutes.
In the generated scripts, mainly syntactical errors occurred (e.g., incorrect usage of variable identifiers, incomplete input definitions, or missing specification of parameters), which could be easily corrected. 
However, the calls to the respective programs to perform the two tasks were correct.

In contrast, the generated extension for \textit{WF4-Grasslands} was of considerably lower quality.
In this case, several syntactic and semantic errors occurred, e.g., the script uses a non-existing library function, no parallelization code is included, and not all requested computations are performed. 
In this state, the domain expert could not resolve the large number of problems within 20 minutes.
However, when interpreting these results, one must remember that the task in this scenario is also significantly more difficult.
Instead of a short task description and specification of a tool to be used, ChatGPT has to design and generate the source code for a complex data analysis procedure containing multiple sub-steps. 
%


\section{Discussion}
We conducted three studies to investigate the capabilities of using ChatGPT for comprehending, modifying, and extending scientific workflows. We discuss our methodology and the results in the following.


\subsection{Comprehending Scientific Workflows}
Study~I was designed to answer \textbf{RQ1} by evaluating ChatGPT's performance in comprehending existing workflows.  
%
%
The domain experts assessed that ChatGPT is good at this task while showing slight differences between the investigated research domains.
In particular, the explanations for workflow \textit{WF4-Grasslands} revealed considerable performance drops.
Unlike the other workflows investigated, this one uses multiple proprietary R and Python scripts instead of leveraging external tools for assembling data processing pipelines.
The lack of standardized tools makes workflow comprehension more challenging since ChatGPT has to interpret complex processing logic and has fewer possibilities to leverage static information, like the description of the general purpose of an established bioinformatics tool, seen through its training while generating the response. 
In addition, code quality and its readability may strongly influence the results for workflows containing proprietary scripts.
For instance, one major problem while explaining \textit{WF4-Grass-lands} in Study~I was the misinterpretation of the abbreviation \textit{"fnf"} as  \textit{“fraction of non-forest”}, instead of \textit{"fold and fill"}. Such customized and ambiguous terms challenge \acp{LLM} and reduce their applicability. 
%
%
%
\begin{table*}[tbhp]
    \centering
    \caption{Overview of the used prompts to investigate ChatGPT's capabilities in extending a given partial workflow (Study III). We distinguish two types of prompts: workflow exploration and workflow extension. For the latter, we developed two variants specially designed for tool- (P3\_2a) and script-based workflows (P3\_2b).} 
    \begin{tabular}{cp{1.5cm}p{14.5cm}}
            \toprule
        \textbf{ID}&\textbf{Category}&\textbf{Prompt}\\
            \midrule
            P3\_1&
            \makecell[{{p{1.25cm}}}]{\centering Workflow\\Exploration}&
            \makecell*[{{p{14.5cm}}}]{%
                The following text contains a \textit{[domain]} workflow written in Nextflow: \\
                \textit{[workflow-description]} \\
                The workflow should be used to \textit{[overall-goal]}. Which steps are missing in order to perform \textit{[overall-goal]}? Please specify only the absolutely necessary steps. For each step name up to three \textit{[domain]} tools that can be used to perform the task.
            }\\
        \hline
            P3\_2a&
            \makecell[{{p{1.25cm}}}]{\centering Workflow\\Extension}&
            \makecell*[{{p{14.5cm}}}]{%
                The following text contains a \textit{[domain]} workflow written in Nextflow: \\
                \textit{[workflow-description]}\\
                Please extend to the given workflow to include one further step which \textit{[step-description]} using \textit{[tool]}. Please specify the new process description in a file at \textit{[file-name]}. Please use version 2 of the Nextflow workflow language. The new process should take the output of \textit{[predecessor-step]} as input.
            }\\
        \hline
            P3\_2b&
            \makecell[{{p{1.25cm}}}]{\centering Workflow\\Extension}&
            \makecell*[{{p{14.5cm}}}]{%
                The following text contains a \textit{[domain]} workflow written in Nextflow: \\
                \textit{[workflow-description]}\\
                Please extend to the given workflow to include one further task which performs \textit{[step]} using an \textit{[programming-language]} script. For this, please generate an \textit{[programming-language]} script, stored in \textit{[script-file-name]}, which performs the following computations:\\
                \textit{[verbal-task-description]}\\
                Next to the \textit{[programming-language]} script  generate the Nextflow process description in a file named \textit{[process-file-name]} and the updated workflow. Please use version 2 of the Nextflow workflow language. The new process should take the output of \textit{[predecessor-step]} as input.
            }\\
        %
            \bottomrule
    \end{tabular}
    \label{tab:prompts_us3}
\end{table*}
\subsection{Modifying Scientific Workflows}
We answer \textbf{RQ2} in Study~II by evaluating the modification performance of ChatGPT.
To this end, we requested the LLM to substitute the leveraged tools for two computational tasks, read quality control and reference genome alignment, in the biomedical workflow \textit{WF2-RS-Star}.
The study results suggest that ChatGPT can effectively explore and explain alternative tools in the field, possibly shortening the time the experts spend searching for suitable replacements on the web. 
In contrast, the results also indicate that ChatGPT rather poorly supports the generation of workflow scripts for using these alternative tools.
In only one scenario, i.e., substituting \textit{FASTP} with \textit{Trimmomatic}, the produced script could be run without syntactical errors, and in one other scenario, i.e., replacing STAR with Bowtie, the script could be fixed within 20 minutes to be syntactically and semantically valid.
In the used version of ChatGPT and the selected setup, an increase in efficiency cannot be recorded or anticipated, highlighting the need for further research efforts. 
However, when interpreting the results, it is essential to remember that ChatGPT is a general-purpose LLM rather focusing on human language. 
A potential option for improvement could be testing generative models more strongly adapted to programming code, such as GitHub Copilot or Code Llama~\cite{touvron2023llama}.

\subsection{Extending Scientific Workflows}
Finally, we investigate \textbf{RQ3} by conducting Study III. 
To this end, we requested ChatGPT to extend an existing (partially given) workflow to achieve specific goals.
The study results confirm the findings from the two previous studies and emphasize ChatGPT's difficulties in solving more complex and exploratory problems.
In this case, explaining the necessary steps to answer the given research questions and the generation of the workflow script for the next step offered (partly) severe issues. 
Similar to the results of Study I, the picture is mixed regarding the different research domains, earth observation and bioinformatics.
For the latter, the generated scripts form a relatively good basis for the implementation, having only (minor) syntactical issues that the expert could quickly fix.
In contrast, in the case of earth observation, the script quality was considerably worse, hindering a fast correction by the expert.
These results imply that efficient user support is possible for pipelines mainly leveraging external tools.
However, further research is necessary to investigate user-support strategies for workflows applying specially implemented analysis scripts.
\subsection{Scientific Workflows in \ac{LLM} Training Data}
\acp{LLM} are trained on large amounts of textual data from the web, including programming code and workflow scripts~\cite{kenton2019bert}.
Therefore, it is crucial to consider whether and to what extent an \ac{LLM} was already able to access the workflow scripts of our study during its pretraining.
According to public information\footnote{\url{https://platform.openai.com/docs/models/gpt-3-5} - \lastaccess}, ChatGPT was trained on data gathered until September 2021, meaning that initial versions of two of the five tested workflows (i.e., \textit{WF3-FORCE2NXF-Rangeland} and \textit{WF5-Force}) could have been part of the LLM's training routine.
However, the specific training dataset used for ChatGPT is not accessible to the public, preventing a conclusive assessment.
To attain a more precise estimate of the potential number of workflow scripts within the training data in general, we initiated searches for scientific workflow repositories on GitHub.
We leverage the repository search engine of the website\footnote{\url{https://github.com/search} - \lastaccess} and use the names of four widely-used workflow management systems, i.e., Apache Airflow, Nextflow, Snakemake, and Taverna as a query term.
We filter all repositories with creation data less than 2021-09-01 from the query results.
Of course, the results must be interpreted carefully since not every repository containing the name has to deal with scientific workflows, even if the names are very peculiar.
Detailed statistics from our search results are in \ref{sec:appendix-search-results}. 
As of 09/2021, there were between 352 and 1,900 repositories containing one of the workflow system names in their description.
Moreover, the results highlight the increasing popularity of workflows since, for all systems except for Taverna, the number of repositories has almost doubled over the last two years. 
We also checked the number of Nextflow pipelines available in nf-core. 
As of September 2021, 35 pipe-lines were published, and 19 were under development\footnote{\url{https://nf-co.re/stats} - \lastaccess}. 
Today, nf-core hosts 55 published pipelines and 33 in development.
In summary, we can hypothesize from these results that ChatGPT can likely rely only on a relatively small base of workflow scripts during its training compared to classical programming code (e.g., GitHub currently hosts over 3.9 million Java and over 2.2 million Python repositories\footnote{Determined by using GitHub repository search and the search queries ``\textit{language:Java}'' and ``\textit{language:Python}''. -- \lastaccess}) making user support for workflow design and implementation particularly challenging.
%
%
\begin{table*}[!t]
  \centering
  \caption{Overview of the results of the workflow modification use case in which the tools performing a specific task are replaced by alternative ones. For each task, we provide the original tool (in parenthesis) and the investigated alternative ones suggested by ChatGPT. For each combination, we highlight (\checkmark=yes,$\times$=no) whether the generated workflow script could be executed (Exec.), whether it is semantically valid (Val.), and whether it could be fixed within 20 minutes (Fix). For the latter, (\checkmark) indicates cases where the script could be fixed to be executable but not entirely semantically correct. Moreover, we provide excerpts from the domain expert's comments.}
    \begin{tabular}{llcccl}
            \toprule
       \multicolumn{1}{c}{\textbf{Task}} &
      \multicolumn{1}{c}{\textbf{Alt. Tool}} &
      \textbf{Exec.} &
      \textbf{Val.} &
      \textbf{Fix} &
      \multicolumn{1}{c}{\textbf{Problems / Comments}}
      \\
    \midrule
    \makecell[l]{Read\\quality\\control\\\textit{(FASTP~\cite{chen2018fastp})}\vspace{1.2cm}}&
      \makecell{Trimmomatic\\\cite{bolger2014trimmomatic}\vspace{2.3cm}}&
      \makecell{\checkmark\vspace{2.7cm}} &
      \makecell{$\times$\vspace{2.7cm}} &
      \makecell{(\checkmark)\vspace{2.7cm}} &
      \makecell[{{p{10.0cm}}}]{%
      $\sbullet$ Missing implementation to perform special quality control feature (i.e. PolyG trimming) \\
      $\sbullet$ Invalid description of workflow differences:\\ 
       - ChatGPT stated that polyG trimming isn't available in Trimmomatic, but it actually is, however extra implementation for data adaptation is needed\\
       - Differences of the generated output files not described accurately}
      \\
      \cline{2-6}
     &
      \makecell{Cutadapt\\\cite{martin2011cutadapt}\vspace{2.2cm}}&
      \makecell{$\times$\vspace{2.6cm}}&
      \makecell{$\times$\vspace{2.6cm}} &
      \makecell{(\checkmark)\vspace{2.6cm}} &
        \makecell[{{p{10.0cm}}}]{%
        $\sbullet$ Wrong program call: syntactically incorrect specification of two parameters\\
        $\sbullet$ Missing provision of valid adapter sequences \\
        $\sbullet$ Invalid description of differences of the workflow script:\\
        - ChatGPT stated that polyG trimming isn't available in Cutadapt, but it actually is\\
        - Differences of the generated output files not described accurately
        }
      \\
    \midrule
    \makecell[{{p{1.55cm}}}]{Genome\\indexing \&\\alignment\\\textit{(STAR~\cite{dobin2013star})}\vspace{0.1cm}} &
      \makecell{HISAT2\\\cite{kim2019graph}\vspace{1.4cm}} &
      \makecell{$\times$\vspace{1.9cm}} &
      \makecell{$\times$\vspace{1.9cm}} &
      \makecell{$\times$\vspace{1.9cm}} &
      \makecell[{{p{10.0cm}}}]{%
        $\sbullet$ Invalid definition and linkage  of in- and output between genome index generation and alignment \\
        $\sbullet$ Syntactically incorrect call of the alignment process \\
         $\sbullet$ Does not take parameter \textit{strandedness} of the input data into account \\ 
         $\sbullet$ Output files aren't generated correctly
        }      
      \\
      \cline{2-6}
     &
      \makecell{Bowtie\\\cite{langmead2009ultrafast}\vspace{0.3cm}} &
      \makecell{$\times$\vspace{0.9cm}} &
      \makecell{$\times$\vspace{0.9cm}} &
      \makecell{\checkmark\vspace{0.9cm}} &
      \makecell[{{p{10.0cm}}}]{%
        $\sbullet$ Invalid definition and linkage  of in- and output between genome index generation and alignment \\
        $\sbullet$ Wrong output definition of genome indexing task
      }
      \\
            \bottomrule
    \end{tabular}%
  \label{tab:us2_results}%
\end{table*}%

\subsection{Prompt Design Challenges}
\label{sec:prompt-design-challenges}
While creating prompts for the studies, we identified several challenges and issues that arose while interacting with ChatGPT.

\subsubsection{Representation of Workflows} For the 
representation of the workflow scripts, there is no straightforward option on how to include them in a prompt. 
The workflow descriptions are often spread over several files containing sub-workflows and task descriptions.
In our approach, we first specify the main workflow and then all sub-workflows and task descriptions in order of occurrence.
However, there might be other, more efficient prompt solutions (with respect to the generative language model).
Furthermore, the workflow scripts might exceed the maximum allowed input length of the language model, e.g., ChatGPT variants allow only for 4K to 16K words / tokens \footnote{\url{https://platform.openai.com/docs/models/gpt-3-5} - \lastaccess} in the input sequence.
In particular, workflows heavily relying on specially implemented scripts having hundreds of code lines will face this issue.
%
%
\subsubsection{Loss of Focus} 
Some of the prompts are very long due to the specification of the entire workflow script, which challenges ChatGPT to maintain focus. 
Adding additional instructions to the prompt helped to avoid or reduce this phenomenon, e.g. for the explanation use case (Study I) we added to the prompt \textit{``Don't explain nextflow concepts''} (see P1\_1 and P1\_2 in Table~\ref{tab:prompts_us1}) and \textit{``Don't explain the workflow itself''} (P1\_3) to prevent ChatGPT to generate outputs describing features of the workflow management system or the complete workflow when requesting input data specification.
%
%
\begin{table*}[tbhp]
  \centering
  \caption{Overview of the results of the workflow extension use case in which we provide ChatGPT a partial workflow and request the LLM to extend it by one further computational step. 
  For each investigated use case, we highlight (\checkmark=yes, $\times$=no) whether the generated workflow script could be executed (Exec.), whether it is semantically valid (Val.), and whether it could be fixed within 20 minutes (Fix). For the latter, (\checkmark) indicates cases where the script could be fixed to be executable but not entirely semantically correct. Moreover, we provide excerpts from the domain expert's comments.}
    \begin{tabular}{llcccl}
            \toprule
       \multicolumn{1}{c}{\textbf{Workflow}} &
      \multicolumn{1}{c}{\textbf{Task/Tool}} &
      \textbf{Exec.} &
      \textbf{Val.} &
      \textbf{Fix} &
      \multicolumn{1}{c}{\textbf{Problems / Comments}}
      \\
    \midrule
    \makecell[l]{WF2\\RS-Star\vspace{1.5cm}}&
      \makecell{Transcript\\quantification/\\\textit{Cufflinks}\\\cite{trapnell2010transcript}\vspace{0.7cm}}&
      \makecell{$\times$\vspace{2.00cm}} &
      \makecell{$\times$\vspace{2.00cm}} &
      \makecell{\checkmark\vspace{2.00cm}} &
      \makecell[{{p{10cm}}}]{%
      $\sbullet$ Syntax errors: process definition for CUFFLINKS declares one input channel but two were specified \\
      $\sbullet$ Input tuple does not match input set cardinality declared by process definition \\
      $\sbullet$ Wrong variable name: \lstinline{sorted\_bam} (wrong) instead of \lstinline{sample\_bam} 
      }
      \\
      \cline{2-6}
            &
            \makecell{Format\\conversion/\\\textit{SAMtools}\\\cite{10.1093/gigascience/giab008}\vspace{0.4cm}}&
            \makecell{$\times$\vspace{1.4cm}} &
            \makecell{$\times$\vspace{1.4cm}} &
            \makecell{\checkmark\vspace{1.4cm}} &
            \makecell[{{p{10cm}}}]{%
                $\sbullet$ Syntactical errors: Usage of wrong variable name (\lstinline{sample\_sam}) \\
                $\sbullet$ Incorrect syntax for connecting the new task to the previous one \lstinline{SAMTOOLS(STAR\_ALIGN.sample_sam)} (wrong) vs. \lstinline{SAMTOOLS(STAR\_ALIGN.out.sample_sam)} (correct) 
                }
                \\
        \midrule
    \makecell[l]{WF4\\Grasslands\vspace{3.0cm}}&
      \makecell{AR\\analysis/\\\textit{R script}\vspace{2.5cm}}&
      \makecell{$\times$\vspace{3.5cm}} &
      \makecell{$\times$\vspace{3.5cm}} &
      \makecell{$\times$\vspace{3.5cm}} &
      \makecell[{{p{10cm}}}]{%
      $\sbullet$ Input to the R script is a path to a directory, not a TIFF file \\
      $\sbullet$ Incorrect use of remotePARTS library: There is no function called \textit{autoTrend} in this package \\
      $\sbullet$  Calculation needs to be parallelized (as specified in the request) \\
      $\sbullet$ Computation should be implemented for four types of inputs, namely: GV, NPV, SOIL, and SHADE. \\
      $\sbullet$  Desired outputs from the AR model needs to be retrieved and written out (missing) \\
      $\sbullet$ Script declares a Conda environment (Python), not R environment. \\
      }\\ 
            \bottomrule
    \end{tabular}%
  \label{tab:us3_results}%
\end{table*}%

\subsubsection{Technological Details} In some cases, adaptation to technological 
details of the specified workflows were necessary. 
For example, the Nextflow system offers two language versions for describing processing pipelines. 
The Nextflow workflows in our study all used the new version of the language. 
However, when extending workflows in Study~III, we had to specify the desired version (see P3\_2 in Table~\ref{tab:prompts_us3}) to get the correct output. 
This observation is surprising since the partially given workflow is already in the respective version. 
Interestingly, this was only necessary for the workflow extension but not for their modification (P2\_4 in Table~\ref{tab:prompts_us2}) in which the phenomena did not occur. 

In summary, the efficient and effective formation of prompts offers a wide range of possible solutions. 
In our study, we identified initial clues and difficulties, but further research is needed to detect further potential for improving the interaction between domain experts and ChatGPT and generative \acp{LLM} in general.

\subsection{Limitations and Future Work}
\label{sec:limitations}
In the following, we highlight the limitations of this work that merit further research.

\subsubsection{Study Design}
\label{sec:discussion-study-design}
In each of our three studies, we created and provided the prompts for testing ChatGPT's capabilities concerning the different use cases and the domain experts only evaluated the outputs of ChatGPT, leading to a rather indirect interaction between the domain scientist and the \ac{LLM}.
An alternative design for the study would be to have the experts interact directly with ChatGPT by developing and refining the prompts independently.
In addition to assessing the capabilities of ChatGPT, this would have the advantage of gaining initial insights into interaction forms and patterns of the different experts with ChatGPT.
Moreover, this would allow for improved customization of the prompts to the particular research domain and the idiosyncratic properties and characteristics of each workflow.
Extended optimization of the prompting strategy by the domain scientist could lead to better results but reduce potential time savings in solving the actual task.
Our study design was motivated by the fact that the experts had strongly limited time budgets for the study.
For example, even for evaluating ChatGPT's outputs in Study~I, the experts already needed up to three hours to accurately check the generated explanations.
A study design that envisages direct interaction involves high efforts in terms of introduction and explanation to ChatGPT and prompting strategies for the domain scientists, thus limiting the scope of research questions that can be investigated.
In addition, the selected study design has the advantage of using the same prompts for the different domains, which contributes to better comparability of the results and eliminates the influence of differences for individual prompt differences.

In our study, we focused solely on ChatGPT as generative language model. However, there are many other general-purpose models available (e.g., PaLM-2 \cite{anil2023palm}, BARD\footnote{\url{https://bard.google.com/} - \lastaccess} or Llama-2~\cite{touvron2023llama}) as well as models more specially designed for programming tasks (e.g., GitHub Copilot, Code Llama~\cite{roziere2023code} or OpenAI Codex\footnote{\url{https://openai.com/blog/openai-codex} - \lastaccess}) publicly available and worth investigating.
Our studies only highlight the results of ChatGPT in the version used (GPT-3.5) but do not claim generalizability for other \acp{LLM}.
Finally, recent research showed that placebo effects can undermine the validity of study results when user expectations are altered through the presence of an AI~\cite{kloft2023ai, kosch2023the}.
In future work and in the case of using \acp{LLM}, placebo conditions must be included to avoid findings that are not a result of increased user expectations towards the capabilities of ChatGPT. 
%
%

\subsubsection{Prompting Strategy} 
Next to other models, the prompts used in our studies also constitute a limiting factor.
We cannot exclude the possibility that other prompts, using a different structure or wordings, may achieve better results for the investigated use cases.
In addition, it should be emphasized that the generated texts are subject to stochastic processes, which can lead to deviations even when reusing the same prompts.

\subsubsection{Limited Number of Domain Experts} 
In the context of our studies, only four domain experts evaluated the outputs of ChatGPT. In some cases, generated explanations were assessed by one person only (e.g., Study~II).
This low number of experts limits the validity and generalizability of the results and offers the risk of subjective bias. 
However, recruitment for such studies is difficult because the number of potential participants is small and they often have strongly limited time budgets, making study design challenging.
Please note that for experts in the field, even ``just'' familiarizing themselves with an unfamiliar workflow is a challenging and time-consuming endeavor.

\subsubsection{Investigated domains and selected workflows} 
Our study explores real-world workflows from the two domains, bioinformatics and earth observation. 
Of course, these only represent part of the full range of workflows in the natural sciences. 
It constitutes an exciting follow-up research question: how suitable ChatGPT and other generative \acp{LLM} are in other research contexts, such as climate research~\cite{kunkel2020potential} and astronomy~\cite{ahmad2022efficient}, and whether it is possible to identify categories or groups of domains which are particularly well (or poorly) supported.
Furthermore, we only examined two workflow systems, Nextflow and Apache Airflow, leaving other alternatives, such as Snakemake, Taverna, and Pegasus, for future work.

\subsubsection{Explored Use Cases} 
This work focused on comprehending, modifying, and extending workflows with ChatGPT.
These use cases represent only a partial scope of user support opportunities and are worth considering and evaluating other use cases. 
For instance, migrating workflows implemented in legacy workflow management systems to more recent ones, e.g., transforming Taverna \cite{oinn2004taverna} scripts to Snakemake or Nextflow, or adapting them to different infrastructure stacks poses an interesting research question. 
Moreover, user support in workflow debugging, error identification, or optimization, as done in classical programming~\cite{vaithilingam2022expectation}, would be a valuable contribution to research scientists.

\section{Conclusion}
The significance of large-scale data analysis workflows in advancing research in the natural sciences is growing steadily.
Developers of such workflows, primarily researchers from diverse scientific fields, are challenged with the increasing complexity and scale of their analyses, which involve (next to their domain knowledge) working with different frameworks, tools, programming languages, and infrastructure stacks.
Although a few tools for creating and maintaining workflows are available, improving user efficiency remains an open research area.
In this work, we contribute to this situation by evaluating the suitability of ChatGPT for comprehending, modifying, and extending scientific workflows. 
In three user studies with four researchers from different scientific domains, we evaluated the correctness of ChatGPT regarding explainability, exchange of software components, and extension when providing real-world scientific workflow descriptions.
Our results show a high accuracy for comprehending and explaining scientific workflows while achieving a reduced performance for modifying and extending workflow descriptions. 
These findings clearly illustrate the need for further research in this area.

\section*{Acknowledgments}
This work is supported by \textit{German Research Foundation (DFG)}, CRC 1404: "FONDA: Foundations of Workflows for Large-Scale Scientific Data Analysis" (Project-ID 414984028).

\section*{Declaration of competing interest}
The authors declare that they have no known competing financial interests or personal relationships that could have appeared
to influence the work reported in this paper.

\section*{Author contributions}
\textbf{Mario Sänger}: Conceptualization, Methodology, Investigation, Visualization,  Writing - Original Draft 
\textbf{Ninon De Mecquenem}: Resources, Data Curation, Writing - Review \& Editing
\textbf{Katarzyna Ewa Lewińska}: Resources, Data Curation, Writing - Review \& Editing
\textbf{Vasilis Bountris}: Resources, Data Curation, Writing - Review \& Editing
\textbf{Fabian Lehmann}: Resources, Data Curation, Writing - Review \& Editing
\textbf{Ulf Leser}: Conceptualization, Writing - Review \& Editing, Project administration, Funding acquisition
\textbf{Thomas Kosch}: Conceptualization, Methodology, Writing - Original Draft , Writing - Review \& Editing, Validation, Project administration

\section*{Declaration of generative AI and AI-assisted technologies in the writing process}

During the preparation of this work the author(s) used Grammarly and ChatGPT for linguistic revision. After using this tool/service, the author(s) reviewed and edited the content as needed and take(s) full responsibility for the content of the publication.

\bibliographystyle{elsarticle-num}
\bibliography{references}
\clearpage

\appendix
\counterwithin{figure}{section}   
\counterwithin{table}{section}   

\onecolumn

\renewcommand{\thetable}{A.\arabic{table}}
\renewcommand{\thefigure}{A.\arabic{figure}}

\section{Questionnaire Study I}
\label{sec:appendix-form-us1}
\begin{table*}[!h]
    \centering
    \caption{Feedback form for the first user study which investigates the capabilities of ChatGPT to capture the content of a workflow description. For each item, we added a comment field to report issues and errors in the generated explanations if the domain expert does not  fully apply the content.}
        \begin{tabular}{ccp{7.5cm}p{7.2cm}}
            \toprule
            \textbf{Prompt}&\textbf{ID}&\textbf{Question}&\textbf{Answer options} \\
                    \midrule
                P1\_1 &
                    Q1\_1 &
                    The generated explanation matches the research area of the workflow &
                    5-point Likert scale \\
                \cline{2-4}
                &
                    Q1\_2 &
                    The generated explanation matches the overall aim of the workflow &
                   5-point Likert scale \\
                \hline
                P1\_2 &
                    Q1\_3 &
                    How many of the tasks of the workflow (see task overview below) are contained in the description? &
                    Numerical \\
                \cline{2-4}
                &
                    Q1\_4 &
                    The explanation describes the tasks of the workflow correctly &
                    \makecell[tl]{
                        (1) Don't know\\
                        (2) None of the tasks is correct\\
                        (3) Few tasks are correct\\
                        (4) Most tasks are correct\\
                        (5) All tasks are correct\\
                    }\\
                \cline{2-4}
                &
                    Q1\_5&
                    How many of the software programs / tools used in the workflow are mentioned in the explanation? &
                    Numerical\\
                \cline{2-4}
                &
                    Q1\_6 &
                    The explanation of the software programs / tools used is correct &
                    \makecell[tl]{
                        (1) Don't know \\
                        (2) None of the software programs / tools  is correct \\
                        (3) Few software programs / tools  are correct \\
                        (4) Most software programs / tools  are correct \\
                        (5) All software programs / tools are correct  
                    } \\
                \hline
                P1\_3 & 
                    Q1\_7 &
                    The explanation matches the input data specification of the workflow &
                    5-point Likert scale \\
                \hline
                P1\_4 &
                    Q1\_8 &
                    The generated explanation matches the overall result of the workflow &
                    5-point Likert scale \\
                \hline
                P1\_5 & 
                    Q1\_9 &
                    How many of the generated research questions are valid? &
                    Numerical\\
                    \bottomrule
        \end{tabular}
    \label{tab:form_us1}
\end{table*}
\clearpage

\renewcommand{\thetable}{B.\arabic{table}}
\renewcommand{\thefigure}{B.\arabic{figure}}

\section{Questionnaire Study II}
\label{sec:appendix-form-us2}
\begin{table*}[htbp]
    \centering
    \caption{Feedback form for the second user study which investigates the capabilities of ChatGPT in exchanging the used tools in a scientific workflow. For each item we added a comment field to report issues and errors in the generated explanations if the domain expert doesn’t fully apply with the content.}
        \begin{tabular}{ccp{11.5cm}p{3cm}}
            \toprule
            \textbf{Prompt}&\textbf{ID}&\textbf{Question}&\textbf{Answer options} \\
                    \midrule
                P2\_1 &
                    Q2\_1 &
                    How many of the 10 alternative tools are valid? &
                    Numerical \\
                \hline
                P2\_2 &
                    Q2\_2 &
                    The selected tools are reasonable alternatives for the task? &
                    5-point Likert scale \\
                \cline{2-4}
                &
                    Q2\_3 &
                    The generated explanation for the first tool highlights the suitability of the tool for the task well &
                    5-point Likert scale\\
                \cline{2-4}
                &
                    Q2\_4 &
                    The generated explanation for second tool highlights the suitability of the tool for the task well &
                    5-point Likert scale\\
                \hline
                P2\_3 & 
                    Q2\_5 &
                    The generated explanation helps to understand the methodical differences &
                    5-point Likert scale \\
                \cline{2-4}
                &
                    Q2\_6 &
                    The generated explanation highlights strengthens and weaknesses of \textit{[original-tool]} correctly&
                    5-point Likert scale \\
                \cline{2-4}
                &
                    Q2\_7 &
                    The generated explanation highlights strengthens and weaknesses of \textit{[alternative-tool]} correctly&
                    5-point Likert scale \\
                \hline
                P2\_4 & 
                    Q2\_8 &
                    The adaption of the workflow is correct  &
                    5-point Likert scale\\
                \cline{2-4}
                &
                    Q2\_9 &
                    The adaption of the task \textit{[task-name]} is correct &
                    5-point Likert scale \\
                \cline{2-4}
                &
                    Q2\_10 &
                    The explanation of the not-supported features by \textit{[alternative-tool]} is correct  &
                    5-point Likert scale \\
                \cline{2-4}
                &
                    Q2\_11 &
                    Can the workflow be executed without errors?  &
                    Yes/No\\
                \cline{2-4}
                &
                    Q2\_12 &
                    How long did it take to correct the workflow? &
                    Numerical\\
                \cline{2-4}
                &
                    Q2\_13 &
                    What had to be adapted to make the workflow executable? &
                    Free-text\\
                \bottomrule
        \end{tabular}
    \label{tab:form_us2}
\end{table*}
\clearpage

\renewcommand{\thetable}{C.\arabic{table}}
\renewcommand{\thefigure}{C.\arabic{figure}}

\section{Questionnaire Study III}
\label{sec:appendix-form-us3}
\begin{table*}[h!]
    \centering
    \caption{Feedback form for the third user study which investigates the capabilities of ChatGPT to extend a given (partial) workflow script. For each item we added a comment field to report issues and errors in the generated explanations if the domain expert doesn’t fully apply with the content.}
        \begin{tabular}{ccp{10cm}p{3cm}}
            \toprule
            \textbf{Prompt}&\textbf{ID}&\textbf{Question}&\textbf{Answer options} \\
                    \midrule
                P3\_1 &
                    Q3\_1 &
                    Is the list of necessary steps complete? &
                    Yes/No \\
                \cline{2-4}
                &
                    Q3\_2 &
                    How many of the proposed steps are correct? Please give your answer in the form of "x out of y steps". &
                   Free text \\
                \cline{2-4}
                &
                    Q3\_3 &
                    How many of the proposed tools are correct? Please give your answer in the form of "x out of y tools". &
                   Free text \\
                \hline
                P3\_2a/b &
                    Q3\_4 &
                    The extension of the workflow is correct&
                    5-point Likert scale \\
                \cline{2-4}
                &
                    Q3\_5 &
                    Can the workflow be executed without errors? &
                    Yes/No\\
                \cline{2-4}
                &
                    Q3\_6&
                    How long did it take to correct the workflow? Enter "not finished" when you set yourself a time limit that has run up. &
                    Free text\\
                \cline{2-4}
                &
                    Q3\_7 &
                    What had to be adapted to make the workflow executable? &
                    Free text\\
                    \bottomrule
        \end{tabular}
    \label{tab:form_us3}
\end{table*}
\clearpage

\renewcommand{\thetable}{D.\arabic{table}}
\renewcommand{\thefigure}{D.\arabic{figure}}

\section{Schema WF4-Grasslands}
\label{sec:appendix-grasslands}
\begin{figure*}[tbhp]
\includegraphics[width=0.8\textwidth]{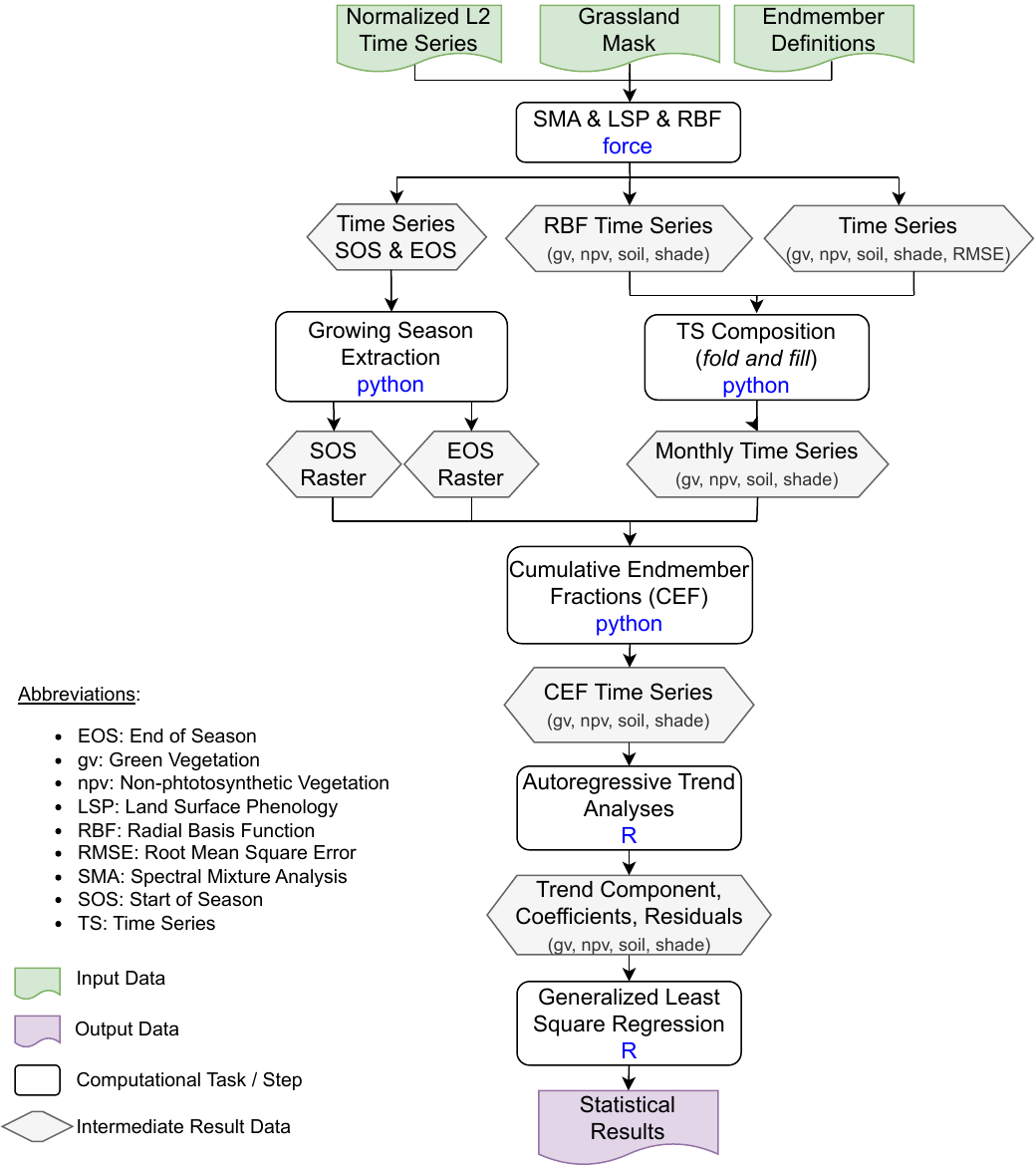}
\centering
\caption{Overview of the earth observation workflow \textit{WF4-Grasslands} developed by one of the domain experts. 
The workflow aims at understanding differences in long-term changes (1984-2022) in ground cover fractions specific to European grasslands depending on the definition of endmembers (i.e., unique spectral signatures of a specific material or ground cover)  approximating these fractions.
The figure highlights the conceptual schema and data flow of the workflow.}
\label{fig:wf4_grasslands}
\end{figure*}
\clearpage

\renewcommand{\thetable}{E.\arabic{table}}
\renewcommand{\thefigure}{E.\arabic{figure}}

\section{Verbal Task Description}
\label{sec:appendix-verbal-description}
In Study III, we use the following task description, provided by the earth observation expert, for extending \textit{WF4-Grasslands} with an autoregressive trend analysis:

\begin{quote}
- step 1: it reads the data from a multi-band TIFF file into a raster brick as defined with the raster package. The input file needs to be selected based on a location specified by a string parameter passed into the script \\ 
- step 2: it converts all 0s in the data brick to 'NA'\\
- step 3: it fits per-pixel time trend with Auto Regressive trend function from the remotePARTS R package. Each raster in a data brick represents one time step in a time series over which the trend is fitted. The result of the trend fitting should be a multi-band raster in a TIFF format that comprises bands with information on trend's slope, intercept, p-value, temporal auto-correlation, and residuals for each time step. \\
- The steps 1-3 should be applied repetitively and independently to four raster datasets available for each location, namely: GV, NPV, SOIL, and SHADE. \\
- The script should be implemented using parallel cluster processing functionality in the raster package and provide control over the number of nodes, and RAM available to the process.
\end{quote}

\renewcommand{\thetable}{F.\arabic{table}}
\renewcommand{\thefigure}{F.\arabic{figure}}

\section{GitHub Search Results}
\label{sec:appendix-search-results}
\begin{table*}[htbp]
  \centering
  \caption{Statistics of the search results for four different scientific workflow systems using the GitHub search engine.
  For each system we use the system name as search term and restrict the result repositories to be created before the date give by the column (group). 
  }
    \begin{tabular}{l|rr|rr}
    \textbf{Search}&
      \multicolumn{2}{c|}{\textbf{$<$ 2021-09-01}} &
      \multicolumn{2}{c}{\textbf{$<$ 2023-09-01}}
      \\
    \textbf{Term} &
      \textbf{\# Repositories} &
      \textbf{\# Pull requests} &
      \textbf{\# Repositories} &
      \textbf{\# Commits}
      \\
    \midrule
    apache airflow &
      1,200 &
      27,000 &
      2,600 &
      85,000
      \\
    nextflow &
      1,700 &
      5,000 &
      3,200 &
      14,000
      \\
    snakemake &
      1,900 &
      5,000 &
      3,400 &
      18,000
      \\
    taverna &
      352 &
      244 &
      472 &
      645
      \\
    \end{tabular}%
  \label{tab:search-results}
\end{table*}%

\clearpage

\twocolumn

\end{document}